\documentclass[numberedappendix,12pt,preprint]{emulateapj} 
\slugcomment{Accepted by the Astrophysical Journal, 2015 March 18}  % comment overwriting the draft version and date

\pdfoutput=1

\usepackage[]{amsmath}
\usepackage{color}
\usepackage{epstopdf}
\usepackage{enumerate}

\newcommand{\simgt}{\;\rlap{\lower 3.5 pt \hbox{$\mathchar \sim$}} \raise
1pt \hbox {$>$}\;}
\newcommand{\simlt}{\;\rlap{\lower 3.5 pt \hbox{$\mathchar \sim$}} \raise
1pt \hbox {$<$}\;}

\newcommand{\be}{\begin{equation}}
\newcommand{\ee}{\end{equation}}

\shorttitle{LF Magnification Bias}
\shortauthors{Mason et al. (2015)}

\usepackage{color}			      % using color package        - for emulateapj
\definecolor{midgray}{gray}{0.4}		% defining gray color        - for emulateapj
\definecolor{orange}{rgb}{1,0.5,0}    %        - for emulateapj

\usepackage[pdftex,colorlinks=true,citecolor=midgray,linkcolor=midgray]{hyperref}        %    - for emulateapj

\newcommand{\BE}{\begin{equation}}
\newcommand{\EE}{\end{equation}}
\newcommand{\BEA}{\begin{eqnarray}}
\newcommand{\EEA}{\end{eqnarray}}
\newcommand{\Eq}[1]{Equation~(\ref{#1})}

\begin{document}


\title{Correcting the $\lowercase{z}\sim8$ Galaxy Luminosity Function for Gravitational Lensing Magnification Bias}


\author{
Charlotte A. Mason$^{1}$,
Tommaso Treu$^{1,2}$,
Kasper B. Schmidt$^{1}$,
Thomas E. Collett$^{3,4}$,
Michele Trenti$^{5,7}$,
Philip J. Marshall$^{6}$, 
Robert Barone-Nugent$^{7}$, 
Larry D. Bradley$^{8}$,
Massimo Stiavelli$^{8}$, and
Stuart Wyithe$^{7}$
}
\affil{$^{1}$ Department of Physics, University of California, Santa Barbara, CA, 93106-9530, USA}
\affil{$^{2}$ Department of Physics and Astronomy, UCLA, Los Angeles, CA, 90095-1547, USA}
\affil{$^{3}$ Institute of Astronomy, University of Cambridge, Madingley Rd, Cambridge, CB3 0HA, UK}
\affil{$^{4}$ Institute for Cosmology and Gravitation, University of Portsmouth, Burnaby Road, Portsmouth, PO1 3FX, UK}
\affil{$^{5}$ Kavli Institute for Cosmology and Institute of Astronomy, University of Cambridge, Madingley Road, Cambridge, CB3 0HA, UK} 
\affil{$^{6}$ Kavli Institute for Particle Astrophysics and Cosmology, Stanford University, 452 Lomita Mall, Stanford, CA 94035, USA}
\affil{$^{7}$ School of Physics, University of Melbourne, Parkville, Victoria, Australia}
\affil{$^{8}$ Space Telescope Science Institute, 3700 San Martin Drive, Baltimore, MD, 21218, USA}
\email{cmason@physics.ucsb.edu}


% ----------------------------------------------------------------------------


\begin{abstract}

We present a Bayesian framework to account for the magnification bias from both strong and weak gravitational lensing in estimates of high-redshift galaxy luminosity functions.  We illustrate our method by estimating the $z\sim8$ UV luminosity function using a sample of 97 Y-band dropouts (Lyman break galaxies) found in the Brightest of Reionizing Galaxies (BoRG) survey and from the literature. We find the luminosity function is well described by a Schechter function with characteristic magnitude of $M^\star = -19.85^{+0.30}_{-0.35}$, faint-end slope of $\alpha = -1.72^{+0.30}_{-0.29}$, and number density of $\log_{10} \Psi^\star [\textrm{Mpc}^{-3}] = -3.00^{+0.23}_{-0.31}$. These parameters are consistent within the uncertainties with those inferred from the same sample without accounting for the magnification bias, demonstrating that the effect is small for current surveys at $z\sim8$, and cannot account for the apparent overdensity of bright galaxies compared to a Schechter function found recently by \citet{Bowler2014b,Bowler2014a} and \citet{Finkelstein2014}. We estimate that the probability of finding a strongly lensed $z\sim8$ source in our sample is in the range $\sim 3-15 \%$ depending on limiting magnitude. We identify one strongly-lensed candidate and three cases of intermediate lensing in BoRG (estimated magnification $\mu>1.4$) in addition to the previously known candidate group-scale strong lens. Using a range of theoretical luminosity functions we conclude that magnification bias will dominate wide field surveys -- such as those planned for the Euclid and WFIRST missions -- especially at $z>10$. Magnification bias will need to be accounted for in order to derive accurate estimates of high-redshift luminosity functions in these surveys and to distinguish between galaxy formation models.
\end{abstract}


% ===========================================================================
\section{Introduction}\label{sec:intro}

Accurate measurements of the rest-frame UV luminosity function (LF) are crucial for
studying the evolution of galaxies at high redshift and reconstructing the
physics and timeline of cosmic reionization. In recent years,
significant progress has been achieved in measuring the LF out to
$z\sim8$ and beyond based on images taken with the Hubble Space
Telescope in the deep legacy fields, the Hubble Frontier Fields and through parallel programs
\citep[e.g.,][]{Bouwens2007,Bouwens2011,Bradley2012,Oesch2012,Robertson2013,McLure2013,Schenker2013,Schmidt2014,Bradley2014,Bouwens2014,Finkelstein2014,Zitrin2015,Coe2015}.

From many of these surveys it appears the LF at $z<6$ is well fit by a \citet{Schechter1976} function with a power-law slope at faint luminosities 
and an exponential drop at the bright end, where it is expected that feedback reduces star-formation in the most massive galaxies \citep{Somerville2012} and dust extinction may reduce the UV flux of galaxies \citep{Cai2014}. The evolution of the LF is expected to be driven by these processes and the evolution of the underlying halo mass function. It is so far unestablished which processes dominate the evolution and whether there are signification changes in the physical conditions of galaxies forming at high redshifts.

Recent studies by \citet{Bowler2014b,Bowler2014a} and \citet{Finkelstein2014} claimed an over-abundance of galaxies at the bright end of the $z\geq 6$ LF when compared to the fit of a \citet{Schechter1976} function, although \citet{Bouwens2014} found no evidence for a departure from a Schechter-like form at $z\sim 4-8$, largely analyzing the same data sets. An over-abundance of bright galaxies may also be apparent in smaller surveys \citep[e.g.,][]{Ono2012,Hathi2012,Finkelstein2013}. If the departure from an exponential cutoff is confirmed by future observations, this may be an indication of the changing astrophysical conditions of high-redshift galaxies. However, another possible explanation is that the LF remains intrinsically with a Schechter form and the over-abundance of bright galaxies is caused by gravitational lensing magnification bias, which has been predicted to be significant for galaxies at $z\geq 8$ \citep{Wyithe2011}.

While it has long been recognized that the gravitational lensing
effect can be exploited in order to probe intrinsically faint galaxies
-- in particular behind massive clusters of galaxies at moderate
redshift -- \citep[e.g.,][]{Franx1997,Ellis2001,Schmidt2014a,Bowler2014a,Zitrin2015,Atek2015,Coe2015}, the effect in blank fields is much less well appreciated.

In fact, gravitational lensing affects all lines of sight, as the
trajectory of every photon in the universe is perturbed by the
inhomogeneous foreground mass distribution. Though the
effect is generally not as strong as in the fields of massive clusters of
galaxies, even so-called blank field surveys are affected by
gravitational lensing (weak, intermediate, or strong). In practice,
owing to the lensing effect, flux-limited surveys include sources that
should be below the sample threshold, but have been magnified into the
sample. Furthermore, gravitational lensing changes the relation between
observed solid angle and cosmic volume with respect to that expected
for a perfectly homogeneous universe. At fixed detector field-of-view
the intrinsic solid angle observed is smaller for magnification
$\mu>1$ and vice versa. This phenomenon is called {\em magnification bias}
\citep[e.g.,][]{Turner1984,Wyithe2000,Wyithe2011} and it can change the shape of the observed
LF.  Thus, it needs to be accounted for in order to derive
accurate intrinsic LFs from flux-limited samples.

The main aim of this paper is to improve the estimation of the
intrinsic UV LF at high redshift by developing a formalism to 
take into account the magnification bias.  Our
new formalism improves on previous work in several ways: we extend the
analytic strong lensing model of \citet{Wyithe2011} to include the
redshift evolution of the deflector population, and we develop a
technique to treat the intermediate lensing regime and introduce a
framework to include weak lensing effects, neither of which have been
systematically accounted for in any previous estimates of the LF.
Furthermore, by providing probability distribution functions for the
magnification of each dropout and empty field, our formalism can be
directly included in any Bayesian LF parameter estimation, thus
allowing for a rigorous derivation of the related uncertainties.

We present two applications of our formalism. The first application is
the interpretation of the $z\sim8$ dropouts found by the
\emph{Brightest of Reionizing Galaxies
Survey}\footnote{\url{http://borg.physics.ucsb.edu}}, 
\citep[hereafter BoRG,][]{Trenti2011}. After estimating the fraction of sources in BoRG that 
are multiply imaged and presenting one strongly-lensed candidate and three candidate systems with
magnification $\mu>1.4$, we use the extended sample presented by
\citet{Schmidt2014} to derive the LF including the
effects of magnification bias. For this we extend the
Bayesian formalism introduced by \citet{Schmidt2014} by including a term
describing the likelihood for magnification of high-redshift sources
for each field, and marginalize over the range of possible
magnifications.

The second application of our formalism is a set of predictions for
the modification of the LF at $8<z\leq 16$, where JWST will detect dropouts \citep{JWST_SSR}, by
using a variety of possible LFs based on theoretical
models \citep[][]{Munoz2012,Behroozi2015} and extrapolations
of lower redshift data \citep{Bouwens2014,Finkelstein2014}. With our formalism we can give a quantitative assessment of how magnification bias will affect future surveys.

%-  Paper structure/recap

The paper is organized as follows. In Section~\ref{sec:data} we briefly describe the BoRG survey and the data used in this paper. In Section~\ref{sec:theory} we introduce the relevant theoretical background for gravitational lensing and magnification bias. In Section~\ref{sec:strong} we develop a semi-analytic framework, based on that in \citet{Wyithe2011} to study the magnification bias due to strong and intermediate gravitational lensing. In Section~\ref{sec:weak} we use the reconstruction of lines-of-sight in cosmological simulation data to investigate weak lensing. The Bayesian inference for the determination of the intrinsic LF is introduced in Section~\ref{sec:LF} and presented in more detail in Appendix~\ref{app:bayes}. The results are presented and discussed in Section~\ref{sec:results}. A brief summary is given in Section~\ref{sec:conc}.

All magnitudes are AB magnitudes and a standard concordance cosmology with $\Omega_m = 0.3$, $\Omega_\Lambda = 0.7$, and $h = 0.7$ is assumed. The Millennium Simulation uses a cosmology with $\Omega_m = 0.25$, $\Omega_\Lambda = 0.75$, and $h = 0.73$, which is used to estimate the weak lensing magnification. We assume the difference between these two cosmologies is negligible for our purposes.

% ===========================================================================
\section{Data}\label{sec:data}

This paper estimates the $z\sim8$ LF using 38 bright Lyman Break galaxies selected from the BoRG survey and 59 fainter dropouts taken from deep legacy fields (in HUDF09 and the WFC3/IR wide area Early Release Science). The BoRG survey is described briefly in Section~\ref{sec:data-borg}, but we refer to \citet{Trenti2011,Trenti2012,Bradley2012} and \citet{Schmidt2014} for further details. The deep legacy data are described by \citet{Bouwens2011}. Additionally, we used data of galaxies with spectroscopically determined velocity dispersions to estimate the velocity dispersion of the foreground BoRG galaxies (described in Section~\ref{sec:data-lens}). In Section~\ref{sec:data-millsim} we give an overview of the simulated data used in the analysis of weak lensing.

% - - - - - - - - - - - - - - - - - - - - - - - - - - - - - - - - - - - - - - 
\subsection{The BoRG Survey}\label{sec:data-borg}

The ongoing BoRG survey is a pure-parallel imaging program with the HST WFC3. The current survey covers $\sim350$ arcmin$^2$ divided into 71 independent fields located randomly on the sky. This reduces cosmic variance below the level of statistical noise \citep{Trenti2008,Bradley2012}. The photometry is in the visual and near-infrared, primarily using the four HST WFC3 filters F606W, F098M, F125W, and F160W (commonly referred to as V-, Y-, J-, and H-bands respectively). The $z\sim8$ BoRG survey consisted mainly of HST programs GO/PAR 11700 and GO/PAR 12572 (PI: Trenti) and includes a small additional number of coordinated parallels from COS-GTO. 53 core BoRG fields are complemented by other archival data including 8 fields from GO/PAR 11702 \citep[PI: Yan,][]{Yan2011} and 10 COS-GTO fields, which used the F600LP-band instead of the F606W-band. The BoRG survey is the largest current survey of Y-band dropouts by solid angle.

The $z\sim8$ galaxy candidates were identified from Y-band dropouts, full details of the selection criteria used to find dropouts are described in \citet{Schmidt2014}. The BoRG survey detected 38 Lyman break galaxy (LBG) candidates at $z\sim8$ with S/N $>5$ in the J-band, of which 10 have S/N $>8$ \citep{Bradley2012,Schmidt2014}. We use the $5\sigma$ sample of objects in this work. Throughout this work we will assume 42\% of the selected BoRG dropouts are contaminants \citep[usually $z\sim2$ interlopers, see e.g.,][]{Hayes2012,Bouwens2013}. This is the fiducial contamination fraction for the BoRG sample and was shown to be robust in the estimation of the LF by \citet{Bradley2012,Schmidt2014}. By definition we cannot determine which specific sources are contaminants without further photometry and spectroscopy, but our rigorous Bayesian method to determine the LF allows us to accurately estimate the LF parameters accounting for the presence of random contaminants \citep{Schmidt2014}.

% - - - - - - - - - - - - - - - - - - - - - - - - - - - - - - - - - - - - - - 
\subsection{Massive Foreground Galaxies Acting as Deflectors}\label{sec:data-lens}

In Section~\ref{sec:strong-identify} we estimate the velocity dispersions of strong lens candidates in the BoRG fields by comparing their photometry with similar early-type galaxies which have both HST photometry and spectroscopically determined velocity dispersions \citep{Treu2005,Belli2014,Belli2014a}. We divided the galaxy samples into three large redshift bins in order to account for the position of the 4000\AA\ break in the filters at higher redshifts.

In the range $z<1$ we used a sample of 165 spheroidal galaxies from \citet{Treu2005} with photometry from the Great Observatories Origins Deep Survey North \citep[GOODS-N,][]{Bundy2005}. For $z>1$ we use a sample of 66 massive quiescent galaxies, presented by \citet{Belli2014,Belli2014a}, which were selected from HST photometric catalogs of objects in the COSMOS, GOODS and Extended Groth Strip (EGS) fields \citep{Grogin2011,Koekemoer2011,Windhorst2011}. We used an aperture correction to rescale observed velocity dispersions, $\sigma_\textrm{obs}$, to $\sigma_e$, the velocity dispersion within one effective radius, $R_e$. 

We follow \citet{Belli2014} and used the model of \citet{VandeSande2013} which proposes a constant rescaling:
   \BE   \label{eqn:strong-identify_sigmaeff-highz}
      \sigma_e = 1.05\sigma_\textrm{obs}
   \EE
For galaxies at $z<1$ (the \citet{Treu2005} sample), we used the model of \citet{Cappellari2006}:
   \BE   \label{eqn:strong-identify_sigmaeff-lowz}
      \sigma_e = \left(\frac{R_e}{R}\right)^{-0.066}\sigma_\textrm{obs}
   \EE
where the slit size, $R$ is the $1\arcsec$ aperture on Keck DEIMOS \citep{Treu2005}. 

The reference photometry used for the individual samples differ. As listed in Table~\ref{tab:strong-mag-sigma} we use HST F606W for galaxies at $z<0.5$, HST F850LP from \citet{Treu2005} (converted to F098M through linear interpolation) for galaxies at $0.5<z<1.0$, and HST F160W for galaxies at $z>1$.

% - - - - - - - - - - - - - - - - - - - - - - - - - - - - - - - - - - - - - - 
\subsection{The Millennium Simulation}\label{sec:data-millsim}

In Section~\ref{sec:weak} we describe our method to generate weak lensing probability density functions (PDFs) by reconstructing simulation data along the line-of-sight to $z\sim8$. Due to the very high redshift of our sources, it was necessary to use simulation data containing halos out to redshifts above 5.

We used 24 $1.4 \times 1.4$ square degree simulated lightcones built by \citet{Henriques2012} from the Millennium Simulation \citep{Springel2005} which contain halos out to $z\sim12$. While the Millennium Simulation contains halos from very high redshift, it has a box length of only 500 Mpc $h^{-1}$. The comoving distance in the universe to $z=1$ is 2390 Mpc $h^{-1}$, so it is necessary to build lightcones with the galaxies correctly distributed in comoving volumes (see \citet{Blaizot2005} and \citet{Kitzbichler2007} for a thorough discussion of generating mock lightcones).

These lightcones were generated using the semi-analytical galaxy formation model of \citet{Guo2011}, and photometric properties were calculated using the stellar population synthesis code by \citet{Maraston2005} which can be applied at high redshift. 

% ===========================================================================
\section{Theoretical Background}\label{sec:theory}

In this section we summarize the relevant theory for the galaxy LF, strong and weak gravitational lensing, and magnification bias.

%= = = = = = = = = = = = = = = = = = = = = = = = = = = = = = = = = = = = = = 
\subsection{Galaxy Luminosity Function}\label{sec:theory-LF}

When a simply parametrized form is needed, we describe the LF by a Schechter function \citep{Schechter1976}:

   \BE  \label{eqn:theory_LF-sch}
       \Psi(L) = \frac{\Psi^\star}{L^\star}\left( \frac{L}{L^\star} \right)^\alpha \exp\left(-\frac{L}{L^\star}\right)
   \EE

where $L^\star$ marks the characteristic break in the LF, $\Psi^\star$ is the characteristic density at that luminosity and $\alpha$ is the power-law exponent slope of the faint end.

% = = = = = = = = = = = = = = = = = = = = = = = = = = = = = = = = = = = = = = 
\subsection{Strong Lensing}\label{sec:theory-strong}

If the line-of-sight to a background source is closely aligned with a massive foreground object, e.g. a cluster or single massive galaxy, gravitational lensing can produce multiple observed images of the source \citep{Schneider1992,SaasFee}. Multiple imaging signifies the regime of strong gravitational lensing lensing.

% - - - - - - - - - - - - - - - - - - - - - - - - - - - - - - - - - - - - - - 
\subsubsection{Singular Isothermal Sphere}\label{sec:theory-strong-sis}

Strong gravitational lenses are commonly modeled as Singular Isothermal Spheres (SIS), which provides a convenient analytic form to describe the mass profiles of massive galaxies \citep[e.g.][and references therein]{Treu2010}. The scale of image separation is characterized by the {\em Einstein radius} of the lens:
   \BE   \label{eqn:theory-strong_ER-SIS}
   \theta_\textrm{ER}(\sigma, z) = 4\pi \frac{D_{ls}}{D_s} \left(\frac{\sigma}{c}\right)^2
   \EE
where $D_{ls}$ and $D_s$ are the angular diameter distances between the lens and source, and from the observer to the source respectively, $\sigma$ is the velocity dispersion of the lens galaxy, and $c$ is the speed of light. Velocity dispersion is the most important property for determining the strength of a strong gravitational lens as it scales with the mass of the dark matter in the system \citep{Turner1984,SaasFee,Treu2010}.

The magnification, $\mu$, due to an SIS lens is given by:
   \BE   \label{eqn:theory-strong_mu}
   \mu = \frac{|\theta|}{|\theta| - \theta_\textrm{ER}}
   \EE 
where $\theta$ is the distance between the lens and the source in the image plane. An SIS lens can produce two images, with the brighter one having magnification $\mu >2$, or one image with magnification $\mu <2$. The case of multiple imaging is referred to here as strong lensing. In this paper we refer to images with $1.4<\mu<2$ as intermediate lensing.

% - - - - - - - - - - - - - - - - - - - - - - - - - - - - - - - - - - - - - - 
\subsubsection{Multiple Image Optical Depth}\label{sec:theory-strong-optdep}

The optical depth $\tau_m$ is the cross-section for a galaxy at redshift $z_S$ to be multiply imaged (i.e. strongly lensed) by a foreground galaxy at $z_L$: it is the fraction of the sky covered by the Einstein radii of all intervening deflectors at redshifts $z_L$. Following standard practice and assuming SIS deflectors, \citet{Wyithe2011} defines it as:

   \BE   \label{eqn:theory-strong_optdep-wyithe}
   \tau_{m} = \int_0^{z_S} dz_L \int d\sigma \; \Phi(\sigma,z_L) \; (1+z_L)^3 \; c \frac{dt}{dz_L} \pi D_L^2 \; \theta_\textrm{ER}^2(\sigma,z_L)
   \EE
where $\Phi(\sigma,z_L)$ is the velocity dispersion function of the deflectors, $D_L$ is the angular diameter distance to $z_L$, and $t$ is time. Without the magnification bias, the optical depth gives the probability of a high-redshift source being multiply imaged.

%= = = = = = = = = = = = = = = = = = = = = = = = = = = = = = = = = = = = = = 
\subsection{Weak Lensing}\label{sec:theory-weak}

Weak gravitational lensing is the deflection of light that causes the magnification and distortion of an observed source, but without producing multiple images. There are no empty lines-of-sight in the universe, so all light traveling to us has been deflected some amount by intervening mass \citep{Hilbert2007}. Whilst it is impossible to determine the exact effect on individual observed sources, it can be done in a statistical sense and is important to quantify this effect for our high-redshift sources.

The lens equation can be constructed for an arbitrary number of lens planes due to an ensemble of deflectors along the line-of-sight \citep{Hilbert2009,McCully2014}. The magnification of a source in a multiplane system is a function of the total convergence and total shear experienced. \citet{Hilbert2009} showed to first order that the total convergence and shear are the sum of the individual contributions from each object along the line-of-sight:
   \BE   \label{eqn:theory-weak_mu}
   \mu = \frac{1}{(1-\sum_i\kappa_i)^2 - |\sum_i\gamma_i|^2}
   \EE 
The convergence, $\kappa_i$, and shear, $\gamma_i$, of each object are determined by the lens model.

%= = = = = = = = = = = = = = = = = = = = = = = = = = = = = = = = = = = = = = 
\subsection{Magnification Bias}
\label{sec:theory-magbias}

The gravitational lensing of a source with luminosity $L$ in a solid angle $\Omega$ of sky has two effects. The observed luminosity is magnified by a factor $\mu$ and sources are now distributed over a magnified solid angle $\mu \Omega$. In a flux-limited sample intrinsically low luminosity sources can be magnified above the survey limit, while the number density of sources can decrease for a given observed solid angle.

Since the faint end of the LF of high-redshift LBG galaxies is so steep, in regions around large low-redshift deflectors we may observe an excess of intrinsically faint high-redshift sources. These effects are known as the {\em magnification bias} and will effect our inferences about the population and LF of high-redshift galaxies.

If it were possible to observe all galaxies in the universe without the magnification bias the probability of a high-redshift galaxy being strongly lensed is purely given by the optical depth, $\tau_m$ (Section~\ref{sec:theory-strong-optdep}). However, magnification of more numerous intrinsically faint sources into our surveys implies that we do not observe the true population of galaxies with luminosity. The magnification bias increases the probability that a sample of observed high-redshift sources have been gravitationally lensed.

The magnification bias for sources with observed luminosities above $L_\textrm{lim}$ in a flux-limited sample is given by:
   \BE   \label{eqn:theory-magbias}
   B = \frac{\int_{\mu_\textrm{min}}^{\mu_\textrm{max}} d\mu \, p(\mu) N\left(>\frac{L_\textrm{lim}}{\mu}\right)}{N(>L_\textrm{lim})}
   \EE
assuming that each source could be magnified between $\mu_\textrm{min}$ and $\mu_\textrm{max}$. Where $p(\mu)$ is the probability distribution for magnification of a source and $N(>L_\textrm{lim})$ is the integrated galaxy LF \citep{Wyithe2011}.

The true probability of a high-redshift source being multiply imaged is $B\tau_m$. Therefore, using $B$ it is possible to find the fraction of galaxies at a given redshift in a flux-limited sample that are multiply-imaged:
   \BE   \label{eqn:theory-magbias_fraclens}
   F_\textrm{mult} = \frac{B\tau_m}{B\tau_m + B'(1-\tau_m)}
   \EE
We assume that $B'$, the bias for galaxies to not be multiply imaged is close to unity.

If the survey limit is brighter than the characteristic apparent magnitude of the observed sample the magnification bias is expected to be large, as a large fraction of the observed sources are likely to be intrinsically fainter sources magnified above the detection threshold of the survey.

We can compute the gravitationally lensed LF, including strong and weak gravitational lensing:
   \BEA  \label{eqn:theory-magbias_mod-LF}
   \Psi_\textrm{mod}(L) &=& (1-\tau_m)\frac{1}{\mu_\textrm{demag}}\Psi\left(\frac{L}{\mu_\textrm{demag}}\right) \nonumber \\
    &+& \; \tau_m \int_0^{\infty} d\mu \; \frac{1}{\mu} p(\mu) \Psi\left(\frac{L}{\mu}\right)
   \EEA
Where $\mu_\textrm{demag} < 1$ is introduced such that the mean magnification over the entire sky is unity \citep{Pei1995,Wyithe2011} and $p(\mu)$ is the full probability density for magnification of a high-redshift source, as above. For a Schechter LF, the gravitationally lensed LF is predicted to exhibit a `kick' in the bright end \citep[e.g.,][]{Wyithe2011} due to a pile-up of brightened galaxies, whereas at the faint end the magnification of flux is balanced by  the loss of number density \citep[for faint-end slope $\alpha \sim -2$,][]{Blandford1992} so there is no distortion, even if many strongly lensed faint sources are observed.

%\clearpage

% ===========================================================================
\section{Strong and Intermediate Lensing}\label{sec:strong}

In this section we compute the probability that the $z\sim8$ dropouts are affected by strong and intermediate lensing. First, in Section~\ref{sec:strong-zevol} we compute the strong lensing
optical depth and the probability that a $z\sim8$ source is multiply imaged by foreground massive elliptical galaxy deflectors. We account for evolution of the deflector population based on the
observed stellar mass function. In Section~\ref{sec:strong-identify} we describe our method to identify
sources in the intermediate lensing regime ($1.4 <\mu < 2$). In order to identify these
sources, we estimate the lensing strength of massive foreground
galaxies based on HST photometry and an empirical calibration of the
\citet{Faber1976} relation.
A candidate strongly lensed dropout in the BoRG fields was presented by \citet{Barone-Nugent2013a}, in this paper one more candidate multiply-imaged dropout ($\mu > 2$) is found, and three dropouts may experience significant intermediate magnification. We detail their properties in Table~\ref{tab:strong-strongish}.

% = = = = = = = = = = = = = = = = = = = = = = = = = = = = = = = = = = = = = = 
\subsection{Strong Lensing by an Evolving Deflector Population}
\label{sec:strong-zevol}

In order to compute the strong lensing optical depth and multiple
image probability, we follow \citet{Wyithe2011} and use a simple SIS
lensing model (see Section~\ref{sec:theory-strong-sis}) with a flat
cosmology. Strong lenses are assumed to be uniformly distributed in
the universe and we can calculate the probability of encountering a
strong lens along the line-of-sight to a high-redshift source, i.e. the
lensing optical depth (see Section~\ref{sec:theory-strong-optdep}). 
By considering the number of galaxies observed above a
certain flux limit we can calculate the magnification bias factor, $B$, from \Eq{eqn:theory-magbias}, assuming a Schechter luminosity
function (\Eq{eqn:theory_LF-sch}). For these calculations we use the
$z\sim8$ LF inferred by \citet{Schmidt2014}, with a
characteristic magnitude of $M^\star = -20.15^{+0.29}_{-0.38}$,
faint-end slope of $\alpha = -1.87^{+0.26}_{-0.26}$, and number
density of $\log_{10} \Psi^\star [\textrm{Mpc}^{-3}] =
-3.24^{+0.25}_{-0.24}$. We marginalize over the entire MCMC chain for
each of the Schechter parameters.

In their calculation of the optical depth \citet{Wyithe2011} used the
local velocity dispersion function as measured by SDSS
\citep{Choi2007}. As most strong lenses occur at $z \simlt
1.5$ \citep{Fassnacht2004,Treu2010}, \citet{Wyithe2011} assumed that the velocity dispersion function does not evolve with redshift for massive galaxies. This is consistent with studies of
the velocity dispersion function out to $z \sim 1$
\citep[e.g.,][]{Chae2010,Bezanson2012}. However, significant galaxy
growth and evolution is observed from $z>1$ as structure forms \citep{VandeSande2013,Belli2014}, and we can improve the accuracy of the model by allowing the parameters of
the velocity dispersion function for massive ellipticals to evolve
with redshift. Introducing redshift evolution is expected to reduce 
the optical depth \citep{Barkana1999}.

\begin{figure*}
\includegraphics[width=0.49\textwidth]{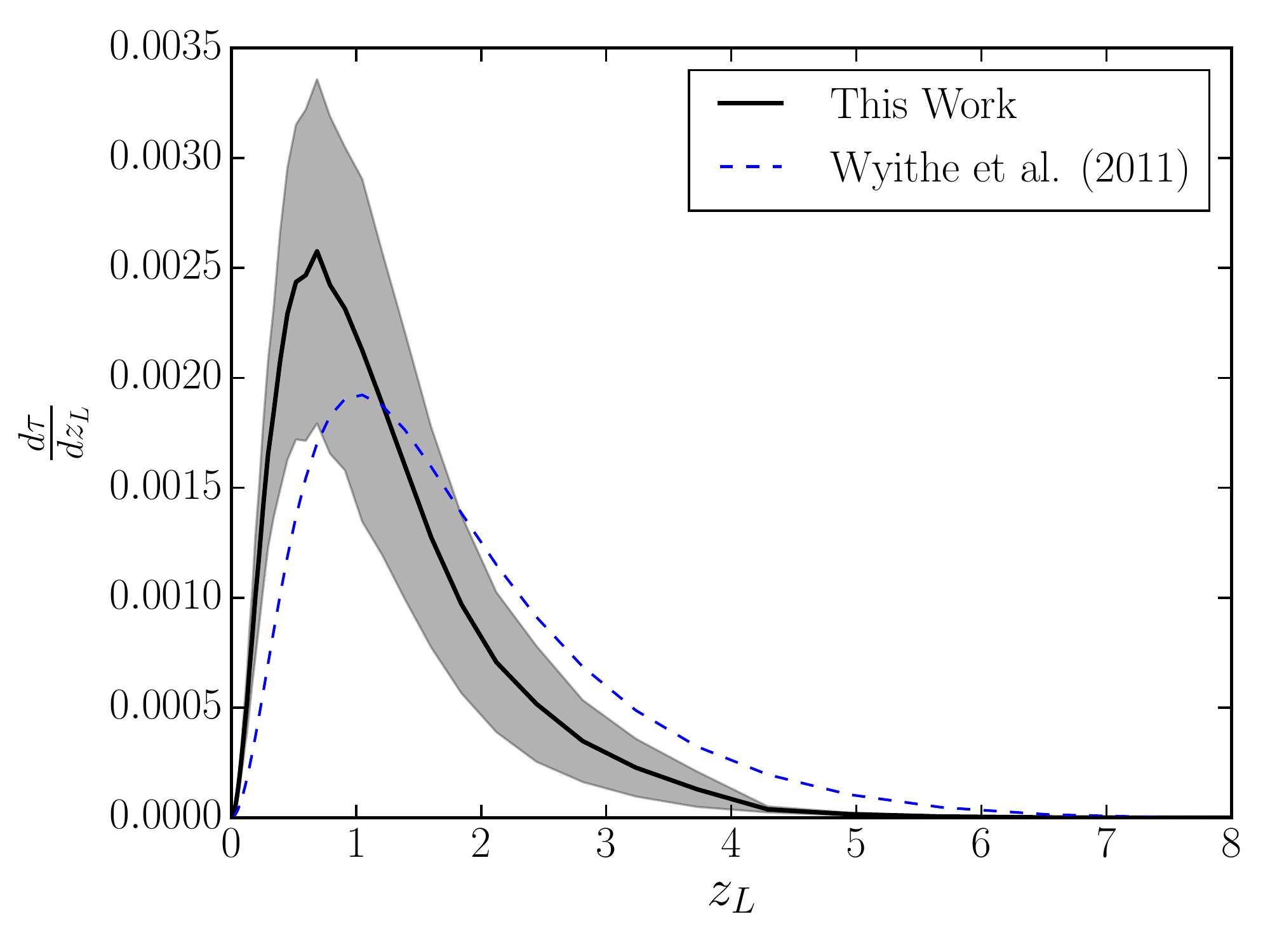}
\includegraphics[width=0.49\textwidth]{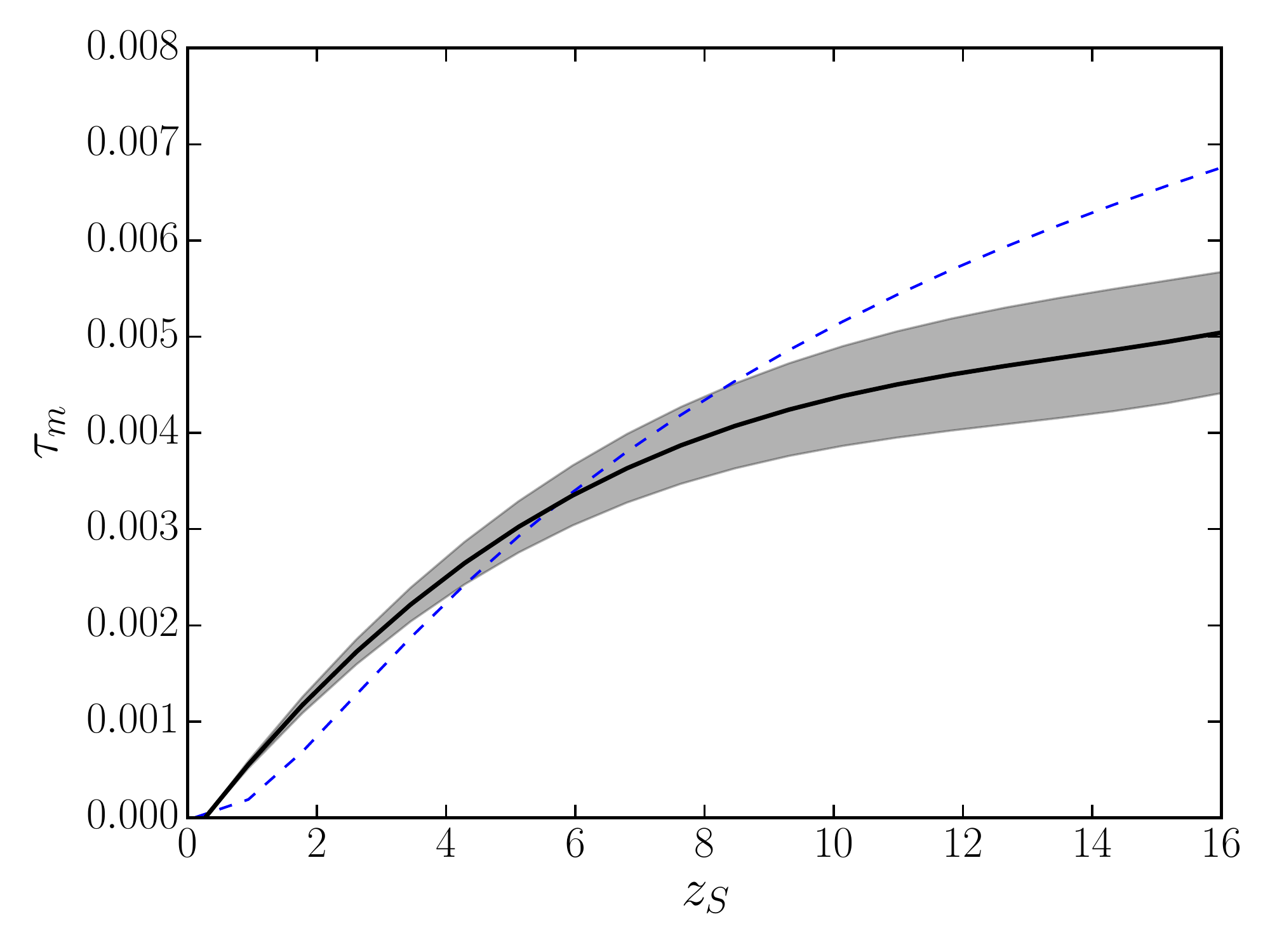}
\caption{{\bf (Left)} Contribution to the optical depth for a source at $z\sim8$ to be multiply imaged as a function of the lens redshift, $z_L$, (solid black line) calculated using \Eq{eqn:theory-strong_optdep-wyithe}, including the evolution of the deflector population with redshift (Section~\ref{sec:strong-zevol}), for comparison we plot the contribution for a constant comoving density of lens galaxies \citep[dashed blue line,][]{Wyithe2011}. {\bf (Right)} Optical depth for multiple imaging as a function of source redshift, including evolution of the deflector population (solid black line). The gray shaded regions show the 1$\sigma$ uncertainty bounds on the optical depth and its distribution, given the uncertainties in velocity dispersion and stellar mass evolution described in the text. The optical depth without redshift evolution of lens galaxies is also plotted for comparison \citep[dashed blue line,][]{Wyithe2011}.}
\label{fig:strong-zevol_optdep-diff-vdfevol}
\end{figure*}

The dashed blue line in the left panel of Figure~\ref{fig:strong-zevol_optdep-diff-vdfevol} shows the probability that the source has been multiply imaged as a function of lens redshift for a source at $z\sim8$, calculated using \Eq{eqn:theory-strong_optdep-wyithe}. The distribution is strongly peaked at $z_L \sim 1$, but there is a significant probability that $z_L > 1.5$. Only 48\% of the contribution to the optical depth for strong lensing occurs at $z_L < 1.5$. We find that 90\% of lensing occurs within a lens redshift of $z_L \simlt 3.5$. Therefore, in order to account for most of the optical depth we need to find the form of the velocity dispersion function out to $z= 3\sim4$ where the galaxy population is significantly different from recent times \citep{Bundy2005,Muzzin2013,VandeSande2014}.

Several studies have investigated the evolution of the velocity dispersion function out to $z \sim 1.5$ \citep[e.g.,][]{Chae2010,Bezanson2011,Bezanson2012,Bezanson2013}. These works are consistent with no evolution, but have large uncertainties. Measurements of velocity dispersion beyond $z > 2$ are very difficult as the brightest emission lines fall within near-IR atmospheric absorption regions \citep{Kriek2006,Belli2014a}.

Therefore, we estimate the evolution of the velocity dispersion function at high redshift based on the evolution of the stellar mass function, a related quantity that has been well-measured at $z>2$. We convert the stellar mass function into the velocity dispersion function by means of the well-known correlation between stellar velocity dispersion ($\sigma$) and stellar mass ($M_\textrm{stell}$) taken from \citet{Auger2010}: $\log(\sigma[\text{km s$^{-1}$}]) = p\overline{M} - 11p + q$, where $p=0.24\pm0.02$, $q=2.34\pm0.01$ and $\overline{M} = \log{(M_\textrm{stell}/M_{\odot})}$. This relation was derived for massive lens galaxies with high velocity dispersions, which will be the strongest contribution to the optical depth as $\tau \sim \sigma^4$.

\begin{figure}[!t] 
\includegraphics[width=0.49\textwidth]{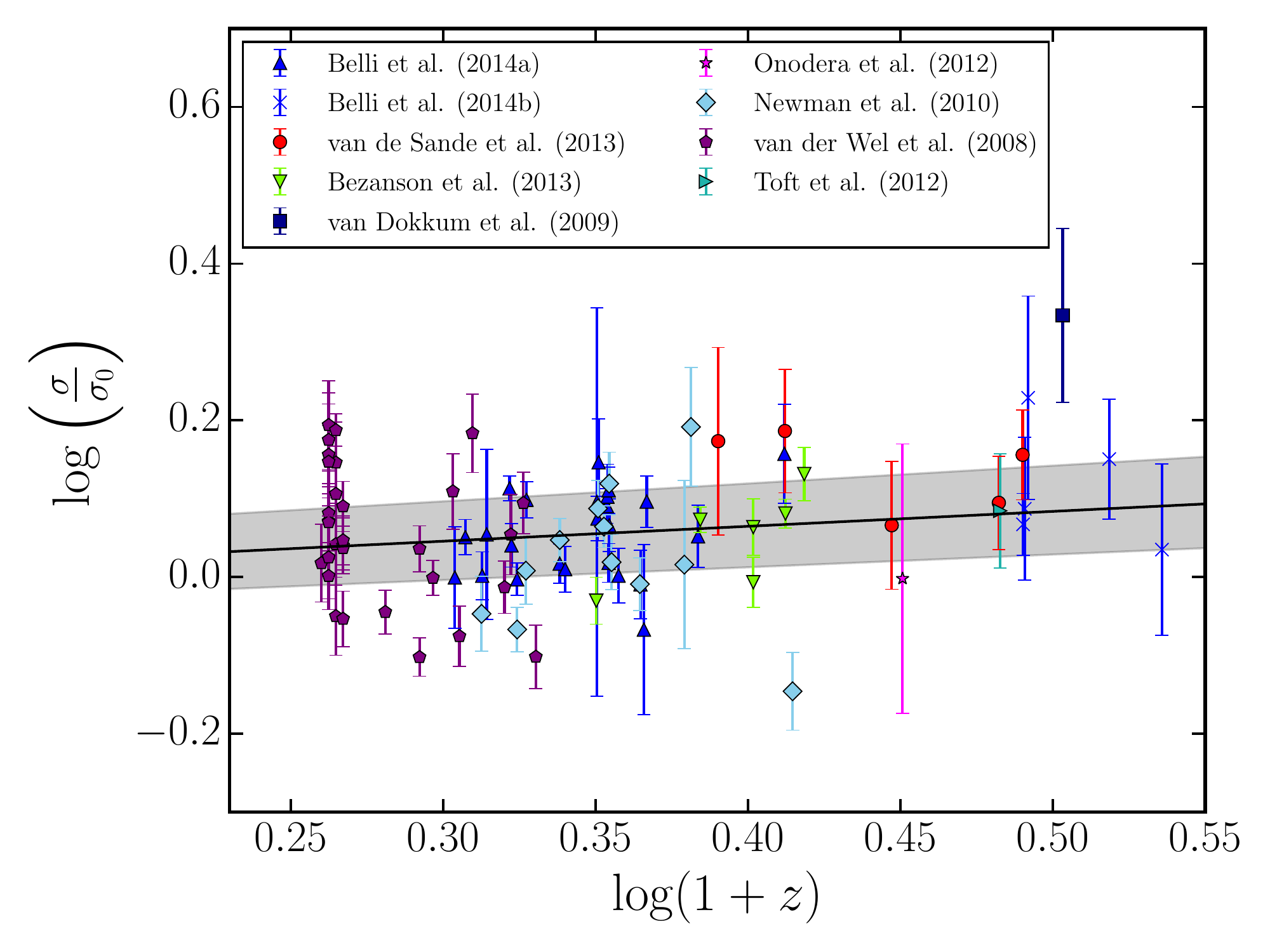}
\caption{Redshift evolution of massive galaxy velocity dispersion, relative to the velocity dispersion estimated from inferred stellar masses via the \citet{Auger2010} relation. We find evolution of the form $(\sigma/\sigma_0) \propto (1+z)^{0.20 \pm 0.07}$, where $\sigma_0$ is the velocity dispersion estimated using the stellar mass-velocity dispersion relation from \citet{Auger2010}. We plot the mean linear fit (black line) and the $1\sigma$ confidence region (gray shaded region).}
\label{fig:strong-zevol_evol-sigma}
\end{figure}

High-redshift galaxies are observed to have higher velocity dispersions at fixed mass than 
in the local universe \citep[e.g.,][]{VandeSande2013,Belli2014,Bezanson2015}. Thus the 
stellar mass-velocity dispersion relation is expected to evolve with redshift. Following 
\citet{VandeSande2013} we expect evolution of the form $(\sigma/\sigma_0) \propto (1+z)^\beta$, where $\sigma_0$ is the expected velocity dispersion at $z\sim0$. 
In Figure~\ref{fig:strong-zevol_evol-sigma} we plot publicly available data from 
\citet{VanderWel2008,vanDokkum2009,Newman2010,Toft2012,Bezanson2013,VandeSande2013,Belli2014,Belli2014a} 
and fit a relation of this form for all galaxies with estimated stellar masses between 
$10.8 < \log(M_\textrm{stell}/M_\odot) < 12.0$, and measured velocity dispersion $\sigma > 200$ km s$^{-1}$ as this was the region where the 
\citet{Auger2010} relation was derived. We find $\beta = 0.20 \pm 0.07$. Our result is lower than the result from 
\citet{VandeSande2013} because we use the \citet{Auger2010} stellar mass-velocity dispersion relation for massive 
lens galaxies as $\sigma_0$, whereas \citet{VandeSande2013} 
compare to a dynamical mass-velocity dispersion relation. As demonstrated in \citet{VandeSande2013} $M_\textrm{stell}/M_\textrm{dyn}$ increases with redshift, so will reduce the evolution we find compared to that in \citet{VandeSande2013}. If we consider the same galaxy sample and fit both our relation derived from stellar masses and the \citet{VandeSande2013} dynamical mass relation, and include the evolution in $M_\textrm{stell}/M_\textrm{dyn}$, our results are consistent.
We note that because the 
optical depth depends on velocity dispersion to the fourth power, the form of the velocity 
dispersion function at $z>2$ is the greatest source of uncertainty in the calculation of optical depths.

% ---------------------------------------------------------------------------
% ---------------------------------------------------------------------------

The stellar mass function can be described by a Schechter function \citep[e.g.,][]{Muzzin2013}:
   \BE   \label{eqn:strong-zevol_SMF}
   \Phi_S(\overline{M}) = (\ln{10})\;\Phi^*_S \;10^{\left(\overline{M}-\overline{M}_S^*\right)(1+\alpha_S)} \; \exp\left[{-10^{\overline{M}-\overline{M}^*_S}}\right]
   \EE
The characteristic stellar mass is given by $\overline{M}_S^* = \log{(M^*_\textrm{stell}/M_{\odot})}$, $\Phi^*_S$ is the characteristic density normalization, and $\alpha_S$ is the low-mass-end slope.

\begin{figure}[!t] 
\includegraphics[width=0.49\textwidth]{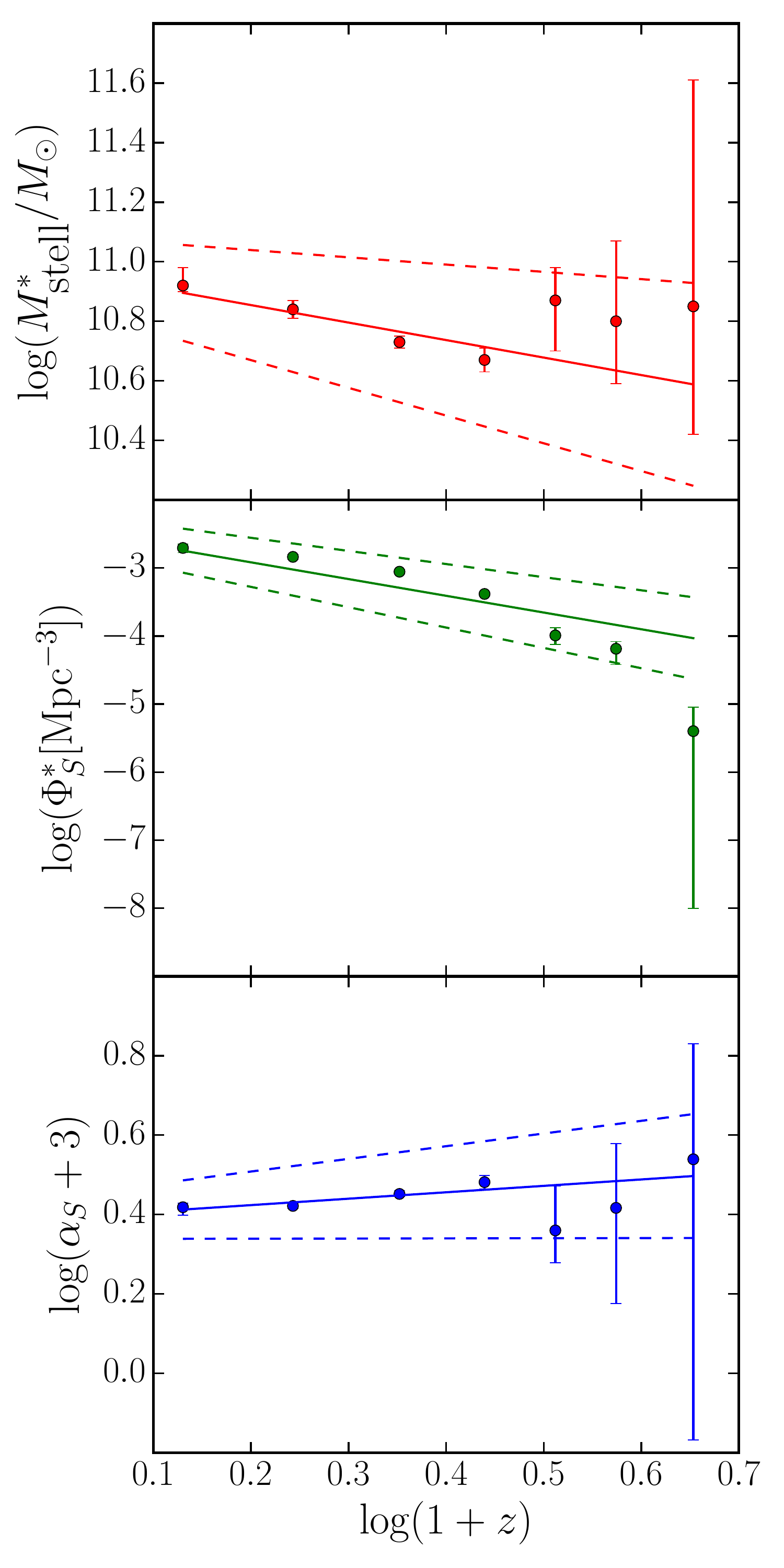}
\caption{Redshift evolution of the best-fit single Schechter function parameters from \citet{Muzzin2013} for the stellar mass function of quiescent galaxies, allowing for evolution of $\alpha_S$. Fits of the form $X_0(1+z)^a$ are plotted: the solid lines show the mean fit, dotted lines show the $1\sigma$ error on the data. Only $\Phi^\star_M$ shows significant evolution with redshift.}
\label{fig:strong-zevol_evol-smf}
\end{figure}

In order to model the redshift evolution of the stellar mass function, we use publicly available data on quiescent galaxies at $z \leq 4$ from the COSMOS/UltraVISTA Survey \citep{Muzzin2013}. They derive the best-fit single Schechter function parameters for the stellar mass function as a function of redshift. Their stellar mass function parameters for quiescent galaxies, allowing for evolution of $\alpha_S$, are plotted as a function of redshift in Figure~\ref{fig:strong-zevol_evol-smf}. We assumed the redshift evolution $X = X_0(1+z)^a$, where $X$ represents the stellar mass function Schechter parameters and $X_0$ represents the values at $z=0$.

We used a Bayesian MCMC linear fitting method to fit this functional form to the data, and plot the mean and one standard deviation confidence fits in Figure~\ref{fig:strong-zevol_evol-smf}. There is significant evolution in $\Phi^*_M$. However, there is also large uncertainty in the evolution of $\Phi^*_M$ due to the spread of the data. We ignore evolution in the low-mass-end slope, since the lensing effect is dominated by the most massive galaxies. We also ignore evolution in $\overline{M}_S^*$, for which the evolution appears non-negligible but it has little effect on \Eq{eqn:strong-zevol_SMF}. The redshift-dependent velocity dispersion function obtained in this way becomes

   \BE   \label{eqn:strong-zevol_vdf-evol}
   \Phi(\sigma,z) = p^{-1} \; \frac{\Phi_S^*(z) }{\sigma (1+z)^\beta} \;\left(\frac{\sigma}{\sigma^*}\right)^{ p^{-1} (1+\alpha_S)} \; \exp\left[{-\left(\frac{\sigma}{\sigma^*}\right)^{ p^{-1} }}\right]
   \EE

with $p=0.24\pm0.02$, $\beta = 0.20\pm0.07$ (obtained from the evolution of velocity dispersion in Figure~\ref{fig:strong-zevol_evol-sigma}), $\Phi_S^*(z) = 3.75 \pm 2.99 \times 10^{-3} (1+z)^{-2.46\pm0.53}$ Mpc$^{-3}$, $\alpha_S=-0.54\pm0.32$ and $\sigma^* = 216\pm18$ km s$^{-1}$. This was derived using the stellar mass-velocity dispersion relation above \citep{Auger2010}, including the scatter in the relation. At $z=0$ recent well-measured velocity dispersion functions \citep[e.g.,][]{Sheth2003,Choi2007} are within the uncertainties of this redshift-evolving relation, showing that our inferred evolution is consistent with direct measurements where they overlap.

% ---------------------------------------------------------------------------
% ---------------------------------------------------------------------------

% ---------------------------------------------------------------------------
% ---------------------------------------------------------------------------

Using this redshift-dependent velocity dispersion function we compute the optical depth for strong lensing, and the distribution of the optical depth with lens redshift. In the left panel of Figure~\ref{fig:strong-zevol_optdep-diff-vdfevol}, now using the redshift evolving deflector population from \Eq{eqn:theory-strong_optdep-wyithe} and \Eq{eqn:strong-zevol_vdf-evol}, we see that the majority of the contribution to the optical depth is from lens galaxies at $z\simlt1.5$, which agrees with current observations of lensed high-redshift dropouts \citep{Barone-Nugent2013a,Schmidt2014a,Atek2015}. In the right panel of Figure~\ref{fig:strong-zevol_optdep-diff-vdfevol} we plot the optical depth as a function of source redshift and find that including the redshift evolution of the deflector population reduces the optical depth at high redshift compared with the work in \citet{Wyithe2011} as expected by theoretical predictions \citep{Barkana1999}, and it appears to start to flatten by $z_S\sim10$. 

Our estimated optical depth at $z<8$ is in good agreement with values derived by an independent method by \citet{Barone-Nugent2015}, and consistent with \citet{Wyithe2011} for $z \simlt 8$. We note the optical depths presented in \cite{Barone-Nugent2015} are marginally higher than the results of this paper, but we can recover their optical depth using a steeper evolution of $\sigma(z)$. It is clear that the uncertainty in the evolution of velocity dispersion, which is the best indicator of the mass of lens galaxies, provides the largest uncertainty in determining the optical depth.

\begin{figure}[!t]
\includegraphics[width=0.49\textwidth]{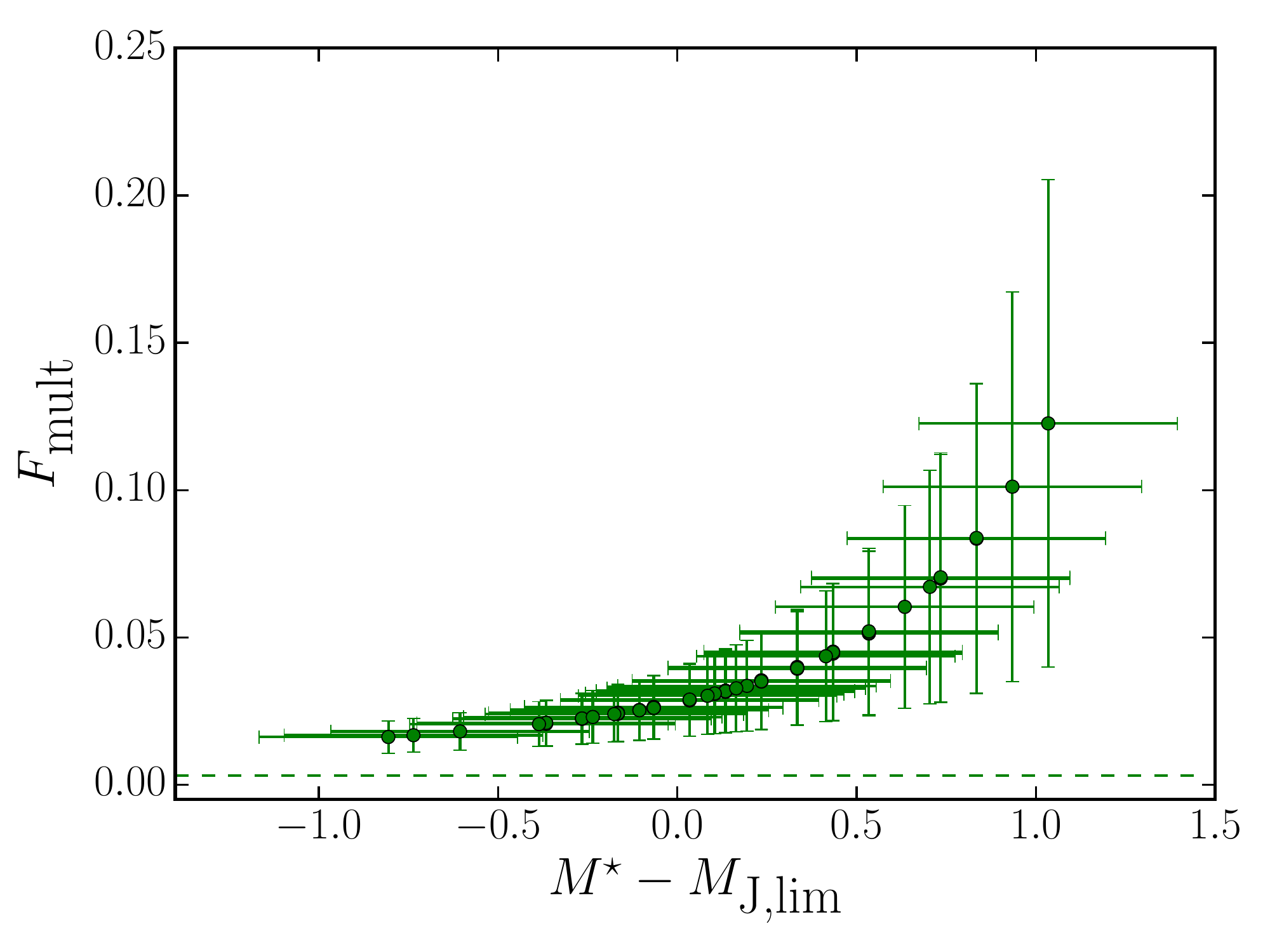}
\caption{Multiply-imaged fraction (see \Eq{eqn:theory-magbias_fraclens}) for $z\sim 8$ sources brighter than the J-band limiting magnitude in each of the BoRG fields, as a function of the UV characteristic magnitude, $M^\star$, including the evolution of the deflector population (Section~\ref{sec:strong-zevol}). The probability of a high-redshift source being multiply imaged increases as the survey magnitude limit becomes brighter than $M^\star$. We expect very few intrinsically bright sources, so any bright source has a high likelihood of being significantly magnified according to the magnification bias. We have used the full MCMC chain for $M^\star$ from \citet{Schmidt2014} and plot the mean value with errorbars of one standard deviation. The optical depth, the probability of multiple imaging without including the magnification bias factor, $B$ (Section~\ref{sec:theory-magbias}), is plotted as the green dashed line.}
\label{fig:strong-zevol_flens}
\end{figure}

Finally, we compute the probability that high-redshift galaxies in a flux-limited sample have been multiply imaged. This is shown in Figure~\ref{fig:strong-zevol_flens} as a function of limiting magnitude for each of the BoRG fields. As expected, the probability that a source in each field is multiply imaged, $F_\textrm{mult}$ (\Eq{eqn:theory-magbias_fraclens}) increases with the survey limiting magnitude, owing to the magnification bias. We estimate 3-15 \% of observed sources brighter than $M^\star$ have been strongly lensed, this is consistent with the results of \citet{Barone-Nugent2015} who use an independent method to infer the lensed fraction.

% ---------------------------------------------------------------------------
% ---------------------------------------------------------------------------

% = = = = = = = = = = = = = = = = = = = = = = = = = = = = = = = = = = = = = = 
\subsection{Identifying Significantly Magnified Sources}\label{sec:strong-identify}

Whilst all the fields are subject to weak lensing, it is necessary to
establish which of the individual sources experience multiple-imaging
($\mu > 2$), or are close enough to a deflector to experience an
intermediate magnification ($1.4 < \mu < 2$). We expect strong lensing evens to be rare, but
possible given the size of the BoRG survey. Among the BoRG sources,
\citet{Barone-Nugent2013a} presented a candidate strongly-lensed system
in borg\_0440-5244 \citep[for naming conventions
see][]{Bradley2012}. The candidate appears to be lensed by a
foreground group with an Einstein radius of $1.49\arcsec$,
corresponding to a velocity dispersion of $\sim 300$ km s$^{-1}$,
producing a magnification of $3.7\pm0.2$ of the dropout.  In this
Section we describe a method to identify other potentially lensed
sources in the catalogs and illustrate how to account for them systematically when estimating the LF.

For computational speed, we considered as potential deflectors only
$z<3$ objects within 18 arcseconds of the $z \sim 8$ dropouts in each
field (the typical Einstein radius is of order 1-2 arcseconds for
massive galaxies). The key quantity that we need to estimate the
lensing strength is the velocity dispersion
\citep{Turner1984,Treu2010}. Thus for every galaxy sufficiently close to a dropout, we estimate their velocity
dispersions by comparing their photometry with that of samples of
similar objects with spectroscopically-determined velocity
dispersions. We selected galaxy samples with HST photometry in bands used in
BoRG in order to estimate velocity dispersion based on our own
photometry. As a comparison sample, we used data from \citet{Treu2005} and
\citet{Belli2014,Belli2014a}, as described in Section~\ref{sec:data-lens}.

As described in Section~\ref{sec:strong-zevol}, the velocity dispersion-stellar mass relation is believed to evolve weakly with redshift since $z\sim2$
\citep[e.g.,][]{VandeSande2013,Belli2014a,Bezanson2015}, and galaxies
will be intrinsically brighter at higher redshift due to younger
stellar populations \citep[e.g.][]{Treu2005}. We account for this by
fitting an evolving \citet{Faber1976} relation to the comparison sample of
the form $L \propto \sigma^4 (1+z)^\beta$.

\begin{figure*}[t]
\centering{
\includegraphics[trim = 0mm 4mm 0mm 20mm, clip, width=0.9\textwidth]{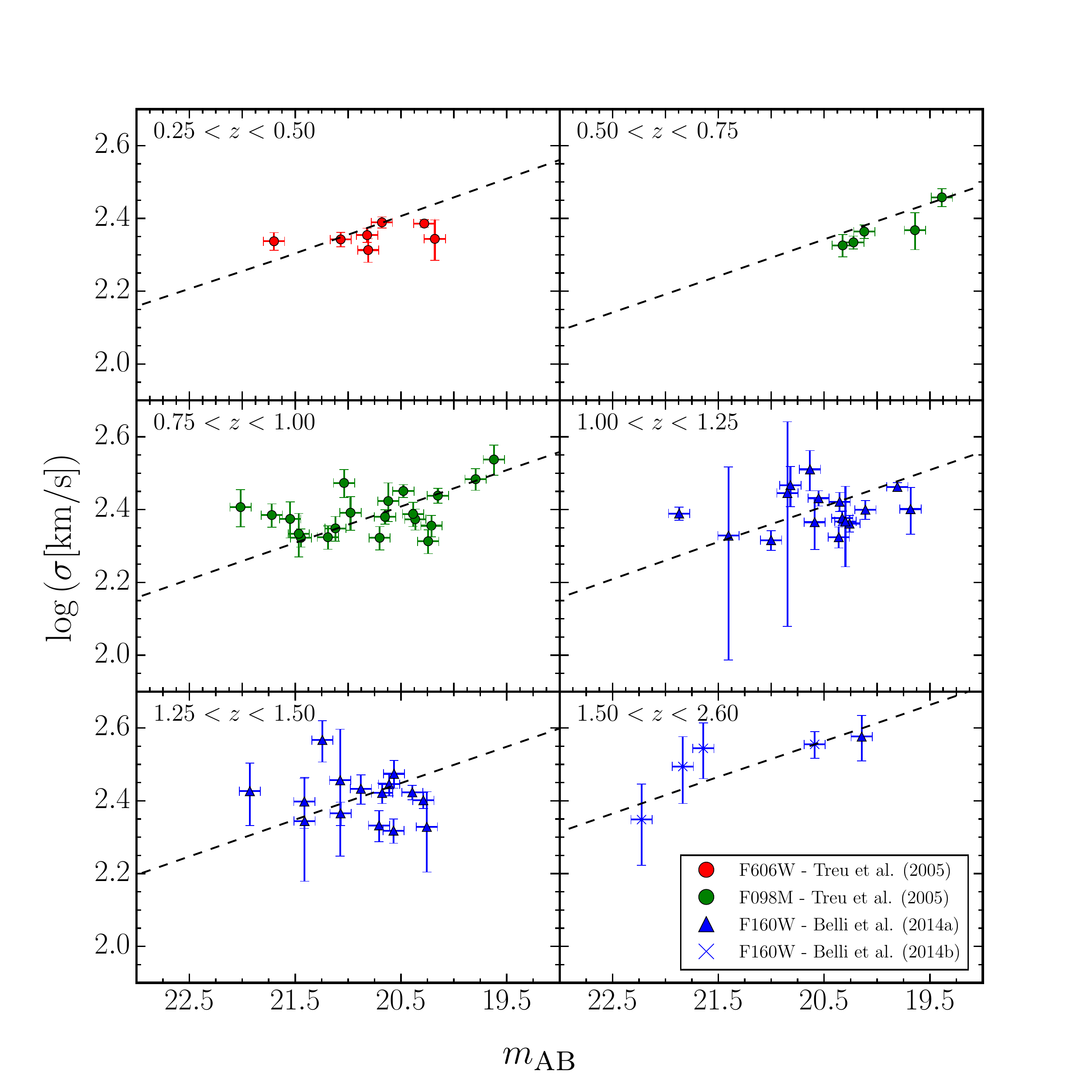}
\caption{Evolving Faber-Jackson relation for massive galaxies with redshift. Data for $z<1$ are from \citet{Treu2005} (red and green circles), data for $z>1$ are from \citet{Belli2014} (blue triangles) and \citet{Belli2014a} (blue crosses). Red points indicate apparent magnitude in the F606W band ($z<0.5$), green points have magnitudes in the F098M band ($0.5<z<1$), and blue points are data with magnitudes in the F160W band ($z>2$). Only galaxies with $\sigma > 200$ km s$^{-1}$ were used in the fitting. The slope of the relation in velocity dispersion and magnitude is fixed at the \citet{Faber1976} result of $L \propto \sigma^4$. We fit the evolution with redshift, which changes the intercept of the line on the velocity dispersion axis. The uncertainty in magnitude is 0.1 mag which is a fiducial value given the fitting procedures. The black dashed lines shows the mean fit for the mean redshift of objects in each plotted bin. The fitting parameters are given in Table~\ref{tab:strong-mag-sigma}.}
\label{fig:strong-identify-FJ}}
\end{figure*}

In practice, we bin the data in redshift, and fit a function of the
form $\log \sigma = -0.1m + a\log(1+z) + b$ using a Bayesian MCMC
estimation where $\sigma$ is the velocity dispersion in km s$^{-1}$, $m$ is
apparent magnitude in a given band, $z$ is galaxy redshift, and $a$
and $b$ are constants. We restrict our fit to galaxies with a measured
velocity dispersion of at least 200 km s$^{-1}$, where samples are
less affected by incompleteness and selection effects. We present the
estimated parameters in Table~\ref{tab:strong-mag-sigma} and fits to
the data are shown in Figure~\ref{fig:strong-identify-FJ}.

%= = = = = = = = = = = = = = = = = = = = = = = = = = = = = = = = = = = = = = = = 
\begin{table}[h]
%\tiny
\centering{
\caption[ ]{Correlation between velocity dispersion, redshift and apparent magnitude}
\label{tab:strong-mag-sigma}
\begin{tabular}[c]{clcc}
\hline
\hline
Redshift     & Band ($m$)   &  a                   & b                     \\
\hline  
$z<0.5$            & F606W   &  $2.26\pm0.79$    &  $4.08\pm0.12$      \\
$0.5 < z < 1.0$    & F089M   &  $0.93\pm0.13$    &  $4.20\pm0.03$      \\
$z>1.0$            & F160W   &  $1.02\pm0.15$    &  $4.12\pm0.05$      \\
\hline
\multicolumn{4}{l}{\textsc{Note.} -- Fits of the form $\log \sigma = -0.1m + a\log(1+z) + b$}
\end{tabular}}
\end{table}
%= = = = = = = = = = = = = = = = = = = = = = = = = = = = = = = = = = = = = = = = 

% ---------------------------------------------------------------------------
% ---------------------------------------------------------------------------

The posterior probability distribution function of Einstein radii for each object are found using \Eq{eqn:theory-strong_ER-SIS}, sampling over the full MCMC chain for the velocity dispersion. The redshifts of the objects were determined using the Bayesian Photometric Redshifts (BPZ) code, using a flat prior and the default parameters and templates \citep{Benitez2004,Coe2006}. All photometric redshifts for relevant foreground galaxies are well-fit by BPZ and have uncertainties in photometric redshift $<15\%$. The PDF for magnification, $p(\mu)$ is found by computing the magnification, $\mu$ (\Eq{eqn:theory-strong_mu}), at the position of the dropout given the distribution of Einstein radii found for each foreground object using the distribution for its velocity dispersion, $\sigma_\textrm{inf}$ estimated from the fits in Table~\ref{tab:strong-mag-sigma}. The greatest source of error in this procedure is the magnitude-velocity dispersion-redshift relation: uncertainties in magnitude and redshift determination have small effects on the magnification PDFs in comparison to the uncertainty in velocity dispersion.

When the mean magnification produced by such a foreground object exceeds $\mu = 1.4$ we use the magnification PDF derived from the above procedure and treat the dropout as described in Section~\ref{sec:LF-combine} in our calculations of the LF.

\begin{figure*}[t]
\centering{
\includegraphics[width=0.8\textwidth]{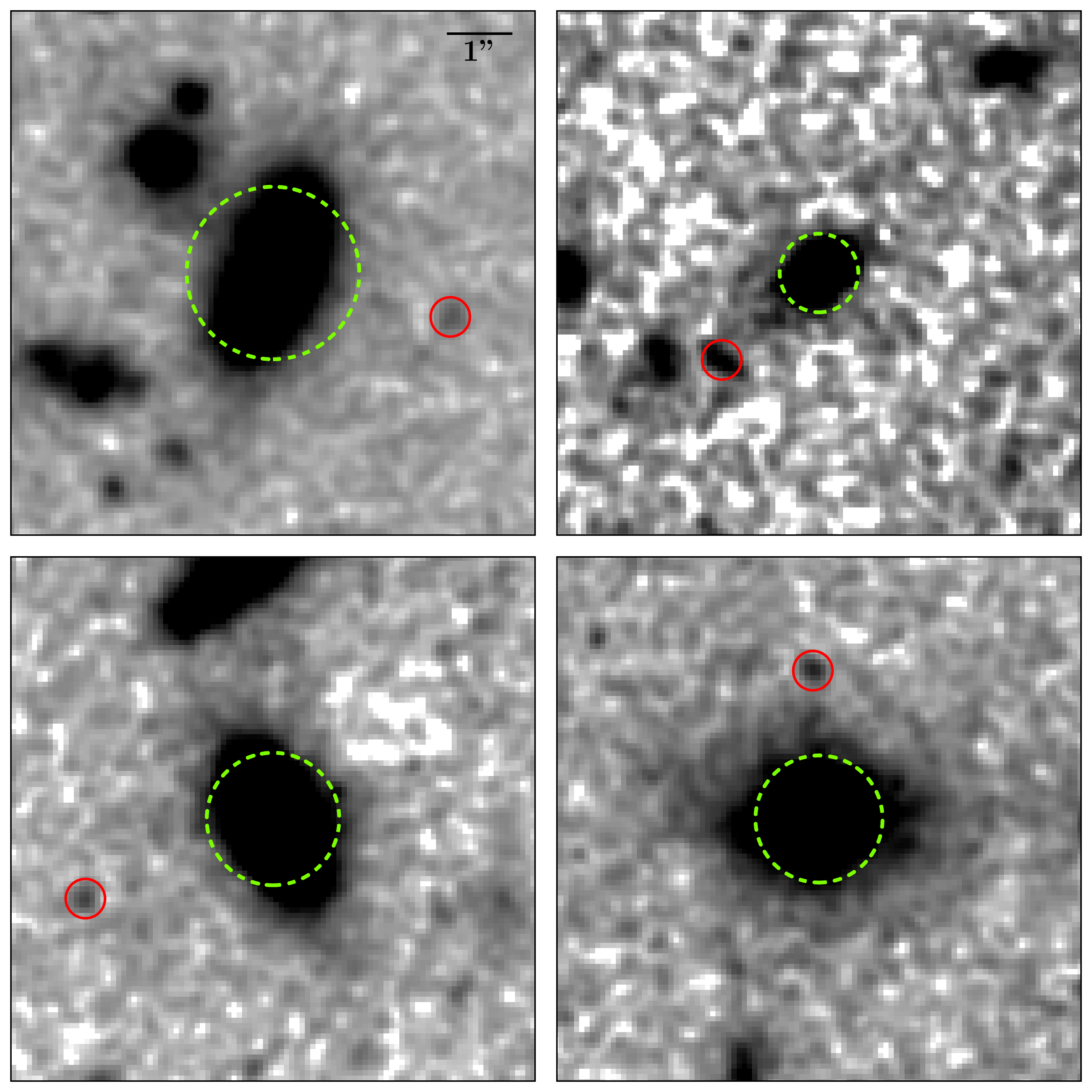}
\caption{The four BoRG dropouts (from top left to bottom right: borg\_0436-5259\_1233, borg\_1301+0000\_160, borg\_1408+5503 and borg\_2155-4411\_341) with significant magnification probabilities, shown in the F125W band with a Gaussian smoothing radius of 1 in $8 \arcsec$ boxes. The solid red lines outline the dropouts with a $0.3 \arcsec$ radius. The dashed green lines outline the potential foreground deflectors, with radius corresponding to the Einstein radius of an SIS deflector lensing a source at $z=8$. The candidate strong lens system (borg\_0436-5259\_1233) is shown in the top left panel, and has an estimated magnification of $\mu = 2.05 \pm 0.52$. Interestingly, borg\_1301+0000\_160 (top right) is the brightest dropout in the BoRG survey. The parameters for all of these objects are given in Table~\ref{tab:strong-strongish}.}
\label{fig:strong-postage}}
\end{figure*}

Using this method, we find one of the dropouts \citep[borg\_0436-5259\_1233, presented in][]{Bradley2012} has a magnification probability distribution consistent with strong lensing. This dropout is shown in the top left panel of Figure~\ref{fig:strong-postage} and its estimated lensing properties are given in Table~\ref{tab:strong-strongish}. The dropout appears to be magnified by a large galaxy at $z\sim 0.40$ with estimated velocity dispersion $294\pm47$ km s$^{-1}$ (estimated from photometry via the empirical relation presented in Table~\ref{tab:strong-mag-sigma}). We estimate its magnification to be $\mu = 2.05 \pm 0.52$, the large uncertainty is due to the uncertainty in the relationship between apparent magnitude and velocity dispersion. If the dropout is indeed strongly lensed the counter image would be almost directly behind the center of the lens galaxy, and will be demagnified according to \Eq{eqn:theory-strong_mu}, unfortunately making it impossible to detect. The dropout is very faint ($m_\textrm{J} = 27.0 \pm 0.2$) and no significant elongation is detected in any of the observed bands but this dropout would be an excellent object for further investigation.

Three of the dropouts (borg\_1301+0000\_160, borg\_1408+5503 and borg\_2155-4411\_341) experience mean
magnification $> 1.4$. Postage stamps of these dropouts are shown in
Figure~\ref{fig:strong-postage} and their lensing
properties are presented in Table~\ref{tab:strong-strongish}. As described in Section~\ref{sec:data-borg} the fiducial BoRG contamination fraction is 42\% \citep{Bradley2012,Schmidt2014} meaning that some of the sources presented here may be lower redshift interlopers \citep[e.g.,][]{Hayes2012,Bouwens2013}. Without further photometry and/or spectroscopy we cannot identify which of the sources are interlopers, but we note that the photometric redshift PDFs for these four sources (obtained from BPZ) all have strong peaks at $z\sim8$, suggesting a higher probability than the average (58\%) for these particular objects to be true $z\sim8$ sources. Interestingly, borg\_1301+0000\_160 is the
brightest dropout in the survey, with $m_J = 25.5\pm0.2$, and appears
tangentially elongated in the J-band image (middle panel of
Figure~\ref{fig:strong-postage}). This object is also a very
interesting target for further imaging and spectroscopic follow-up.

We note that our method assigns a significantly lower velocity
dispersion to the potential strong lens (borg\_0440-5244\_647) than the one estimated by
\citet{Barone-Nugent2013a} in their presentation of this object. 
They estimated the velocity dispersion of
the deflector to be $\sigma \sim 300$ km s$^{-1}$, whereas our method
estimates a mean velocity dispersion of $\sim 170 \pm 33$ km s$^{-1}$. This
is likely to be because our method does not account for lensing by
groups and clusters, while \citet{Barone-Nugent2013a} suggest that this
dropout is lensed by a group of at least two objects at $z\sim1.8$, of
which borg\_0440-5244\_647 is the largest. They estimated velocity
dispersions of the deflector galaxies by using an abundance matching
relation between mass and luminosity, derived from
\citet{Cooray2005}, and measuring the angular size of the lensing
objects. However, when using a redshift-dependent \cite{Faber1976} relation \citep{Barone-Nugent2015} similar to ours (Table~\ref{tab:strong-mag-sigma}) they estimate the velocity dispersion of this single galaxy to be $\sim 180 \pm 46$ km s$^{-1}$ (via private communication), 
which agrees with our result. Neglecting group-scale lensing is a potential limitation of our method, which may
underestimate magnification in a few cases. However the impact on the
overall estimation of the LF inference is negligible
since the phenomenon is so rare.

% ---------------------------------------------------------------------------
% ---------------------------------------------------------------------------

%= = = = = = = = = = = = = = = = = = = = = = = = = = = = = = = = = = = = = = = = 
\begin{table*}[!t]
%\tiny
\centering{
\caption[ ]{Strong and intermediate lensing parameters derived by estimating velocity dispersions of bright foreground galaxies close to $z\sim8$ dropouts}
\label{tab:strong-strongish}
\begin{tabular}[c]{lcccccccc}
\hline
\hline
Field & Dropout ID & $J_\textrm{125} \,^\textrm{a}$  & Foreground ID  & $z_f$  & Separation ($\arcsec$) & $\sigma_\textrm{inf}$(km s$^{-1}$) &  $\theta_\textrm{ER}$ ($\arcsec$) & $\mu$ \\
\hline  
borg\_0436-5259 &  1233$^\textrm{b,c}$ &   $27.1\pm0.2$  &  1191  &  $1.52\pm0.03$     & 2.79   & $294\pm47$  &  $1.32\pm0.40$ &   $2.05\pm0.52$    \\
borg\_1301+0000 &  160$^\textrm{d}$  &   $25.5\pm0.2$  &   144   &  $1.14\pm0.15$     & 1.99   & $184\pm31$  &  $0.60\pm0.20$  &   $1.47\pm0.30$    \\
borg\_1408+5503 &  980$^\textrm{c}$	 &	 $27.0\pm0.2$  &   959   &  $0.40\pm0.06$ 	& 3.11    & $193\pm69$  &  $1.01\pm0.70$  &   $1.54\pm0.62$    \\
borg\_2155-4411 &  341$^\textrm{c}$  &   $26.6\pm0.2$  &  244   &  $0.74\pm0.11$    & 2.27   & $216\pm22$  &  $0.97\pm0.20$  &   $1.80\pm0.33$    \\
\hline
\multicolumn{9}{l}{\textsc{Note.} -- $^\textrm{a}$ Total (AUTOMAG) apparent magnitude in the J-band of the dropout \citep{Bradley2012}.} \\
\multicolumn{9}{l}{ $^\textrm{b}$ Strongly-lensed candidate. $^\textrm{c}$ 5$\sigma$ source. $^\textrm{d}$ 8$\sigma$ source.} 
\end{tabular}}
\end{table*}
%= = = = = = = = = = = = = = = = = = = = = = = = = = = = = = = = = = = = = = = = 
\newpage
% ===========================================================================
\section{Weak Lensing}\label{sec:weak}

In this section we discuss the methods used to find the PDFs for magnification of a source at $z\sim8$ by all intervening matter. We used the Pangloss code\footnote{\url{http://github.com/drphilmarshall/Pangloss}} developed by \citet{Collett2013} that generates lensing parameters for reconstructed lines-of-sight. We describe the production of magnification PDFs from simulation data from the Millennium Simulation \citep{Springel2005} in Section~\ref{sec:weak-pangloss} and in Section~\ref{sec:weak-borg} we present the BoRG field weak magnification PDFs. Our PDFs agree well with other theoretical work at lower redshifts \citep{Hilbert2007,Hilbert2009,Greene2013}.

% = = = = = = = = = = = = = = = = = = = = = = = = = = = = = = = = = = = = = = 
\subsection{Estimating Magnification from Simulation Catalogs}\label{sec:weak-pangloss}

The weak lensing reconstruction model developed by \citet{Collett2013} takes simulation halo catalogs and places halos in a three-dimensional grid, with each halo contributing convergence $\kappa_i$ and shear $\gamma_i$ along a line-of-sight to a source at a given redshift. Halos are modeled as truncated NFW profiles \citep{Baltz2009}:
   \BE   \label{eqn:weak_NFW_trunc}
   \rho(r) = \frac{\rho_\textrm{NFW}(r)}{1 + \left(\frac{r}{r_t}\right)^2}
   \EE
where we used the truncation radius $r_t = 5r_{200}$, shown to be robust by \citet{Collett2013}. Where $r_{200}$ is the radius at which the mass density falls to 200 times the critical mass density of the universe. The convergence and shear derived from this profile are given in \citet{Baltz2009}. Magnification due to all intervening deflectors along a line-of-sight is given by \Eq{eqn:theory-weak_mu}.

We built PDFs for all lensing parameters by sampling over $10^3$ of lines-of-sight. As described in Section~\ref{sec:data-millsim} we used lightcones built from the Millennium Simulation \citep{Henriques2012,Springel2005}.

The simulated catalogs provide a list of halos with associated
galaxies, but they do not include other dark structure, clumped in
filaments and absent in voids. This missing matter will affect the
overall density of the universe so it is necessary to take this into
account when estimating $\kappa$ and $\mu$. We account for this by
subtracting convergence from redshift slices so that the mean
convergence along all lines-of-sight in the catalogs to a given
redshift equals zero, and the mean magnification is unity, as they
should be.

Following work by \citet{Suyu2010} and \citet{Greene2013}, we compare
lines-of-sight in the BoRG fields with simulation data based on
relative density of objects. We define the overdensity parameter
   \BE   \label{eqn:density}
   \xi = \frac{n_i}{n_\textrm{tot}}
   \EE
where $n_i$ is the number of objects per unit area in each lightcone
(or real field) and $n_\textrm{tot}$ is the total number of objects divided
by the total survey area. Given that the simulation catalogs are $\sim
500 \times$ larger than the total BoRG survey area we expect them to
give representative results.

\begin{figure}[t] 
\includegraphics[width=0.49\textwidth]{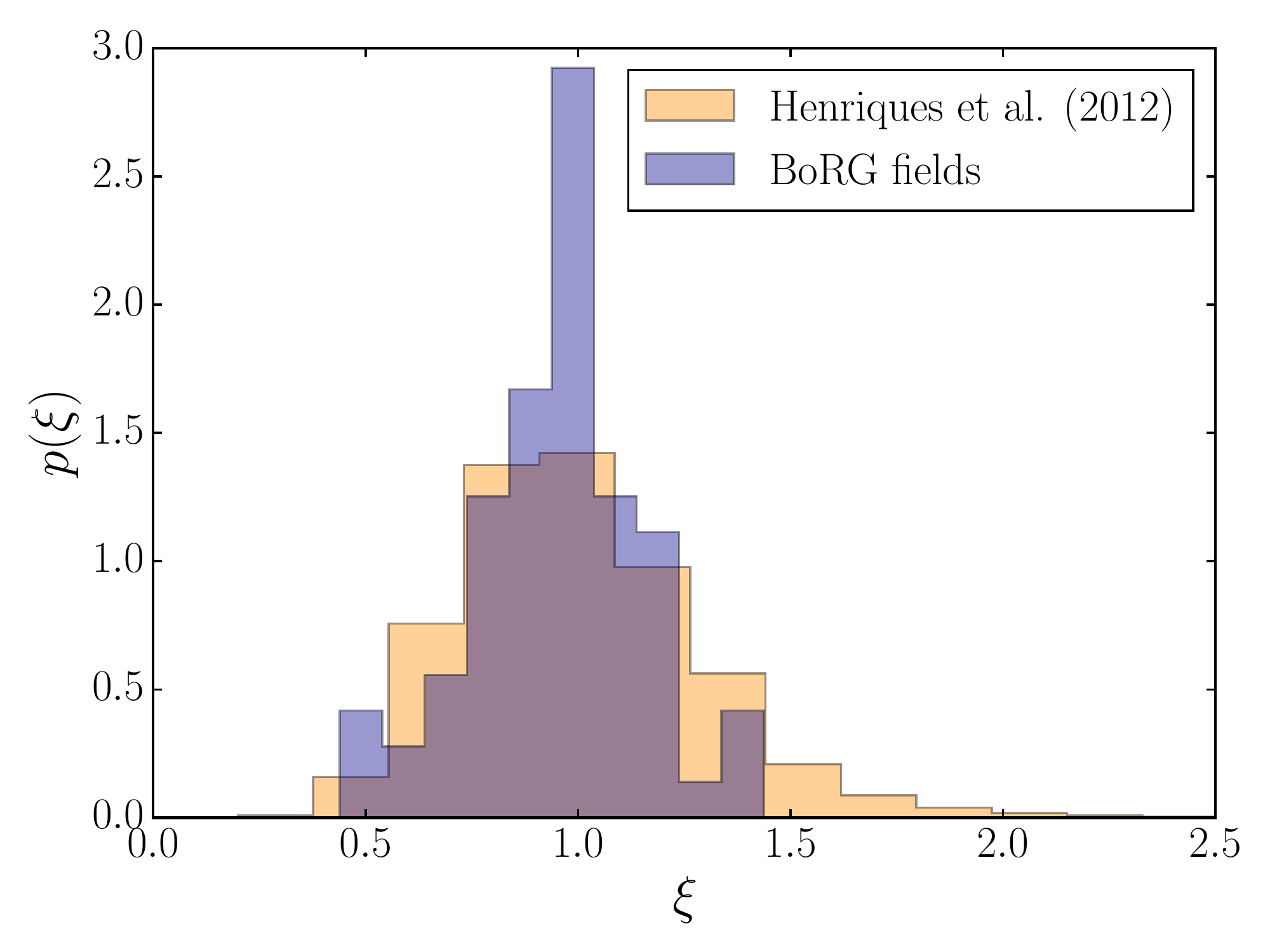}
\caption{Comparison of the overdensity of lines-of-sight in the Millennium Simulation and the BoRG fields. $\xi = n_i/n_\textrm{tot}$ where $n_i$ is the number of objects per unit area above a certain flux limit in each lightcone (or real BoRG field) and $n_\textrm{tot}$ is the total number of objects above the same flux limit divided by the total survey area. We use a flux limit of $m<24$ in F125W (J-band).}
\label{fig:weak-pangloss-millsim_overdensity}
\end{figure}

We then calculate the number of objects per square arcsecond brighter
than $m = 24$ in the J-band in each of the BoRG fields compared to
the total number of objects above this flux limit in the whole
survey. Similarly we calculate the overdensity of objects above the
same limit in the simulated lightcones. \citet{Henriques2012} include
mock photometry based on stellar population synthesis codes by
\citet{Maraston2005} which include J-band magnitudes. As shown in Figure~\ref{fig:weak-pangloss-millsim_overdensity}, the
distribution of overdensities for the observed data is within the
range of that for simulated data. Finally, to generate magnification
PDFs for a given BoRG field, we combine the magnifications from all
simulation lines-of-sight which are within $\pm 2\%$ in overdensity of
the observed value.

% ---------------------------------------------------------------------------
% ---------------------------------------------------------------------------

\begin{figure}[t] 
\includegraphics[width=0.49\textwidth]{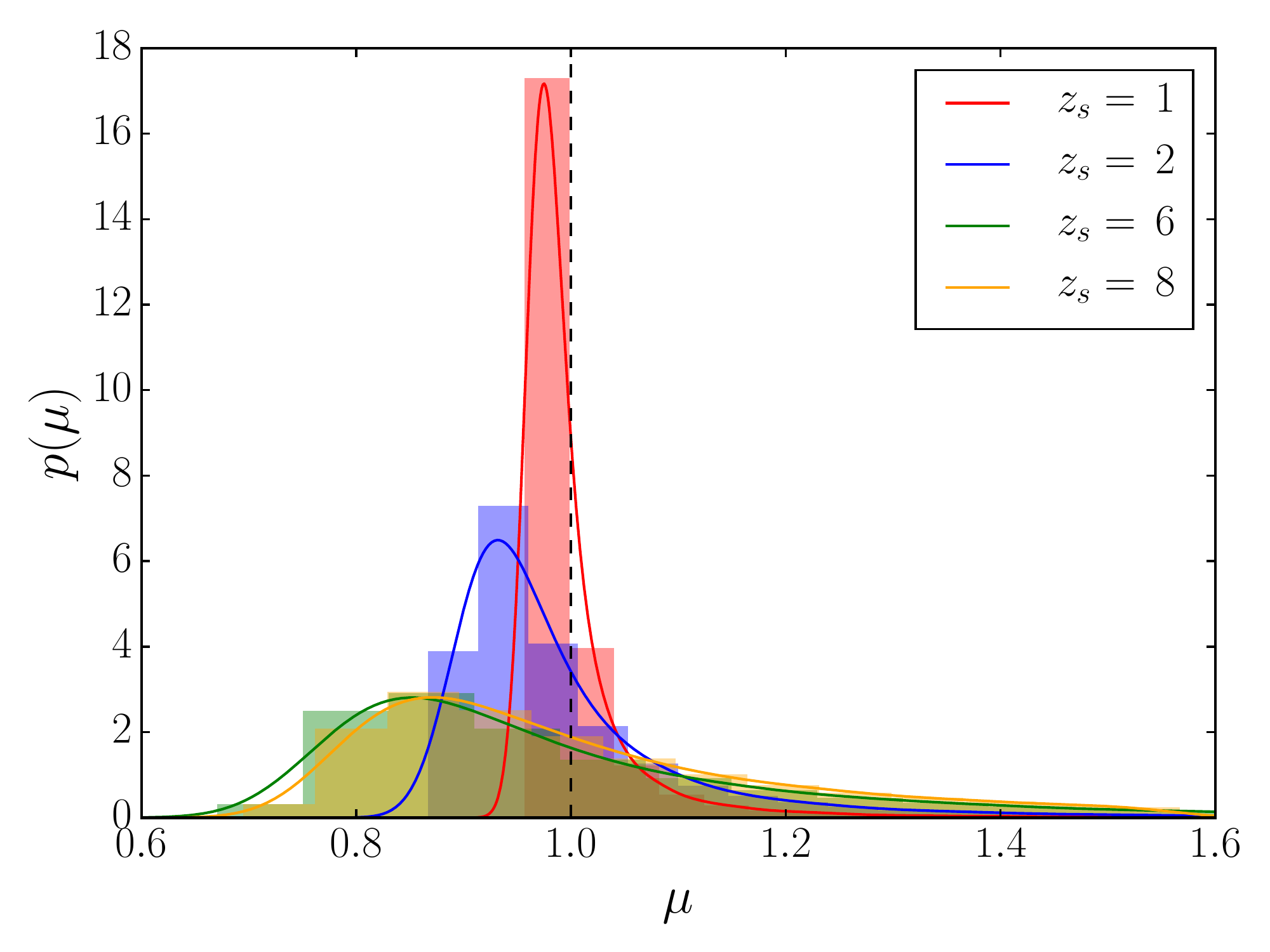}
\caption{Probability distribution function for magnification for four values of source redshift. The dashed line marks the mean magnification of the universe. These results compare well with \citet{Hilbert2007}. Due to the lack of significant mass between $z \sim 6$ and $z \sim 8$ there is little change in the distributions of magnification for sources at those redshifts, as the total convergence does not change much.}
\label{fig:weak-pangloss-compare_Hilbert}
\end{figure}

In Figure~\ref{fig:weak-pangloss-compare_Hilbert} we plot the magnification PDFs for a source at various redshifts over all lines-of-sight. As the source redshift increases, the peak of the distribution shifts to lower magnification, but the high-magnification tail becomes more important, such that the mean magnification over all lines-of-sight remains unity. We match results for $z<6$ from \citet{Hilbert2007} well. It is clear that there is little change in the distribution between $z_S=6$ and $z_S=8$, as there are negligible numbers of large halos above $z>5$.

% ---------------------------------------------------------------------------
% ---------------------------------------------------------------------------

\begin{figure}[t] 
\includegraphics[width=0.49\textwidth]{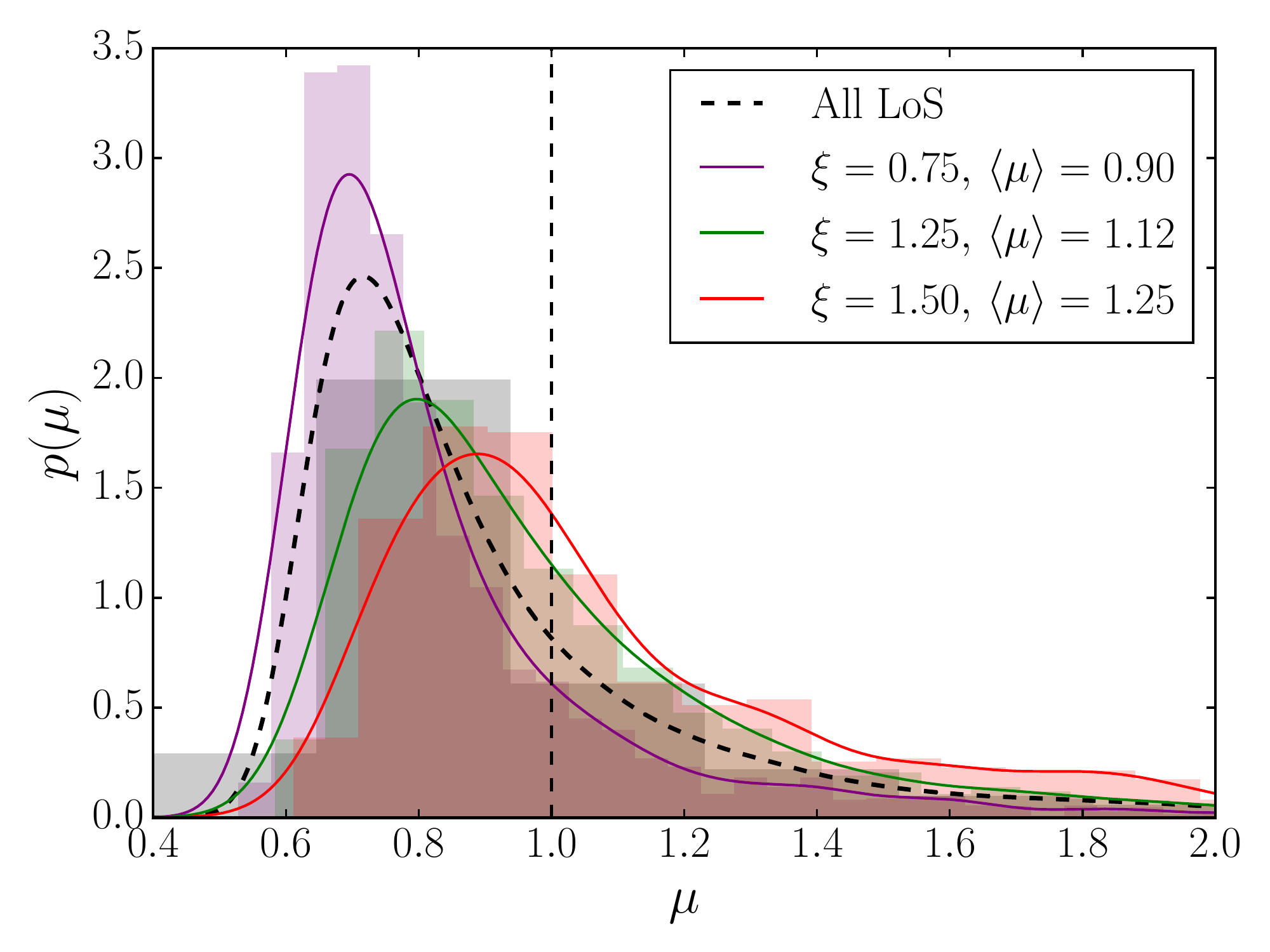}
\caption{Probability distribution function for magnification for a range of values of overdensities for a source at $z=8$. More overdense lines-of-sight are skewed towards higher magnification, with a broad distribution. More underdense lines-of-sight are skewed towards lower magnification, with a narrower distribution due to the deficit of intervening mass.}
\label{fig:weak-pangloss-z8}
\end{figure}

In Figure~\ref{fig:weak-pangloss-z8} we plot the magnification PDFs for a variety of overdensities. The more overdense lines-of-sight produce a higher mean magnification, as expected, but also have a greater variance than the distributions for underdense lines-of-sight. This agrees well with the estimates at lower redshift by \citet{Greene2013}.

% ---------------------------------------------------------------------------
% ---------------------------------------------------------------------------

%= = = = = = = = = = = = = = = = = = = = = = = = = = = = = = = = = = = = = = 
\subsection{BoRG Weak Lensing Magnification PDFs}\label{sec:weak-borg}

\begin{figure}[t] 
\includegraphics[width=0.49\textwidth]{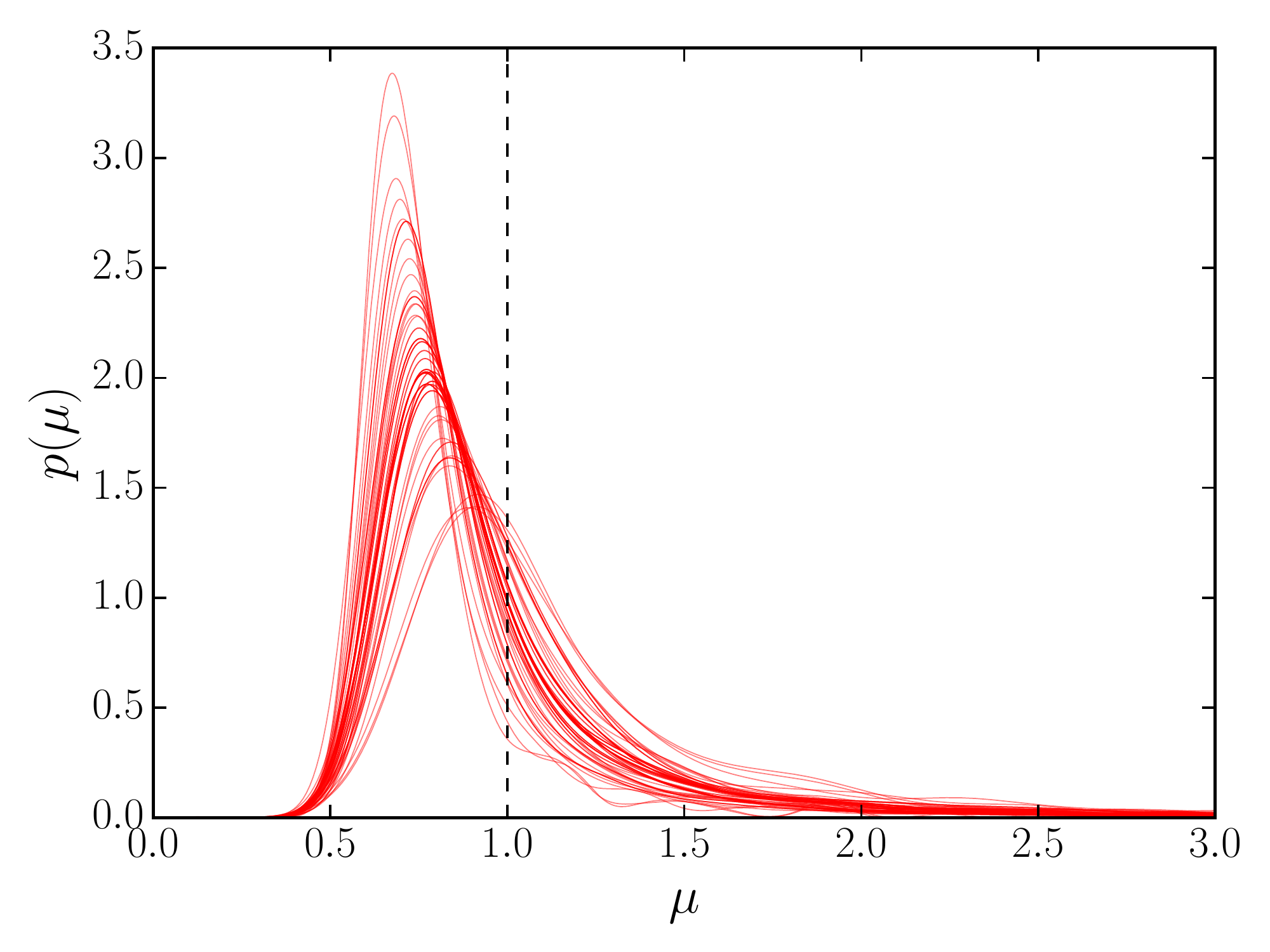}
\caption{Probability distribution function for magnification for all of the BoRG fields, with a source at $z\sim8$. The lines are kernel density estimations to the distributions. It is clear there is little range in overdensity for the BoRG fields.}
\label{fig:weak-borg_allpdfs}
\end{figure}

The kernel density estimates \citep{rosenblatt1956,parzen1962} fit to the magnification PDFs for all the BoRG fields are shown in Figure~\ref{fig:weak-borg_allpdfs}. As expected, the
BoRG fields do not have significant over- or underdensities, but are
rather typical of blank fields at $z\sim8$, as shown in Figure~\ref{fig:weak-pangloss-compare_Hilbert}.

% ---------------------------------------------------------------------------
% ---------------------------------------------------------------------------

There is significant motivation for the magnification PDFs to take a log-normal form. The 3D matter density distribution of the universe is well-described by a log-normal random field \citep{Coles1991}, and weak lensing probability distributions arise directly from the mass distribution. However, when accounting for the magnification bias in individual fields to infer the LF from the dropout sample (see Section~\ref{sec:LF}) it was necessary to express the magnification distributions in a form that could easily convolve analytically with a Gaussian distribution (for more details see Appendix~\ref{app:bayes}). For this we used a Bayesian MCMC approach to fit the distributions of magnification for each field as a linear sum of Gaussian functions.

% ===========================================================================
\section{Recomputing the LF}\label{sec:LF}

In this section we outline the method of estimating the $z\sim8$
LF from the BoRG high-redshift candidates, taking
the magnification bias into account.

Following \citet{Schmidt2014}, who did not account for
the magnification bias when estimating the BoRG $z\sim 8 $ LF, we use the Bayesian inference method devised by
\citet{Kelly2008}, which is described in Section~\ref{sec:LF-bayesian}
and in Appendix~\ref{app:bayes}. In Section~\ref{sec:LF-lensing} we
describe in more detail how we take into account the weak and
intermediate lensing magnification.

%= = = = = = = = = = = = = = = = = = = = = = = = = = = = = = = = = = = = = = 
\subsection{Bayesian Estimation of the LF}
\label{sec:LF-bayesian}

As in \citet{Schmidt2014}, we assume that the intrinsic luminosity
function is modeled by the Schechter function in \Eq{eqn:theory_LF-sch}. In order to
facilitate comparison with our previous work we use the sample
of 38 BoRG Y-band dropouts and 59 additional fainter dropouts
from the Hubble Ultra-Deep Field (HUDF) and Early Release Science (ERS)
programs \citep{Bouwens2011}.

Bayesian statistics allows us to express the posterior probability that the LF is fit by a Schechter function with parameters $\theta = (\alpha, L^\star, \Psi^\star)$ given the observed luminosity $L_\textrm{J,obs}$ of the dropouts in the J-band, and the non-detections in the V-band ($I_V = 0$), as the product of the prior on the Schechter parameters and the likelihood:
   \BE  \label{eqn:LF-bayesian_bayes}
      p(\theta \,|\, L_\textrm{J,obs}, I_V =0) \propto p(\theta) \times p(L_\textrm{J,obs}, I_V = 0 \,|\, \theta)
   \EE
This posterior probability can be expressed (see Appendix~\ref{app:bayes} for full details) as:

   \BEA  \label{eqn:LF-bayesian_bayes-margpost}
      p(\theta &\,|\,& L_\textrm{J,obs}, \; I_\textrm{V}=0) \propto \; p(\theta) \times C^{N_z}_{(1-f)n}  \times C^{\frac{f}{1-f} N_z}_{f n}  \nonumber  \\
      &\times& \; \prod_{l}^\mathcal{C} \left[1-\frac{ A_l}{\overline{\mu}_l A_\textrm{sky}}\; p(I=1\,|\,\theta) \right]^{\frac{N_z-(1-f_l)c_{l}}{1-f_l}}  \nonumber \\
      &\times& \; \prod_i^{n} p(L_{\textrm{J,obs},i}\,|\,\theta)
   \EEA

Where we iterate over $l$ fields with $i$ $z\sim8$ candidates. Here $N_z$ is the number of high-$z$ dropouts in the surveyed comoving cosmological volume, $A_l$ is the area of the individual $\mathcal{C}$ fields in \citet{Schmidt2014}, which each contain $c_l$ high redshift candidates ($n = \sum_l^\mathcal{C} c_l$). Each candidate has an assumed contamination fraction of $f_l$. We use a fiducial value for the contamination of 42\% for the BoRG sample, the contamination fractions for the HUDF/ERS samples are included in the selection function, see Appendix~\ref{app:bayes}, as described in \citet{Oesch2012,Bradley2012} and \citet{Schmidt2014}. Changing the contamination value in the range $f=0-0.60$ effects the characteristic magnitude and the number density of the LF by less than their estimated $1\sigma$ uncertainties, and the change in the faint-end slope is comparable to its $1\sigma$ uncertainty \citep{Bradley2012,Schmidt2014}. The Bayesian framework allows us to accurately estimate the LF parameters accounting for contamination. $A_\textrm{sky}$ is the area of the full sky The $C^a_b$ factors are binomial coefficients which are the fully correct method of modeling source counts.

We assume uniform priors on $\alpha$, $\log_{10}L^\star$ and $\log_{10}N_z$. $p(I=1\,|\,\theta)$ is the probability distribution of an object making it into the dropout sample based on the photometric selection described in \citet{Schmidt2014}. $p(L_{\textrm{J,obs},i}\,|\,\theta)$ is the likelihood function for the observed J-band luminosity of the $i$'th object in the sample.  

The last term includes marginalization over the magnification PDF:
  \BEA \label{eqn:LF-bayesian_bayes-Lobs-like2}
      p(L_\textrm{J,obs}\,|\,\theta) = \int \int& p(\mu)\; p(L_\textrm{J,obs}\,|\,\mu L_\textrm{J,true})  \nonumber \\
     \times &p(L_\textrm{J,true}\,|\,\theta) \; dL_\textrm{J,true} \; d\mu
   \EEA
In Appendix~\ref{app:bayes} we give the expanded expression of the
posterior distribution from \Eq{eqn:LF-bayesian_bayes-margpost} used when performing the LF parameter inference
and describe the derivation and motivation for
\Eq{eqn:LF-bayesian_bayes-Lobs-like2}. 
We refer to Appendix~\ref{app:bayes} and \citet{Schmidt2014} for
further details.

%= = = = = = = = = = = = = = = = = = = = = = = = = = = = = = = = = = = = = = 
\subsection{Including the Lensing Corrections}\label{sec:LF-lensing}

%= = = = = = = = = = = = = = = = = = = = = = = = = = = = = = = = = = = = = = 
\subsubsection{Analytic Form for Magnification PDFs}\label{sec:LF-bayesian-pofmu}

In order to make integration of \Eq{eqn:LF-bayesian_bayes-Lobs-like2} computationally feasible we require a simple analytic form for $p(\mu)$ that will convolve simply with a Gaussian distribution (see Appendix~\ref{app:bayes}). As described in Section~\ref{sec:weak-borg}, the weak lensing magnification PDF is well-fit by a log-normal distribution. However, this cannot be convolved analytically with a Gaussian.

Therefore, we fit the magnification PDFs from all regimes as a linear combination of Gaussian functions with different means and standard deviations. The weak lensing magnification PDFs (see Section~\ref{sec:weak-borg}) are well-fit by a combination of three Gaussian functions. The intermediate lensing PDFs (see Section~\ref{sec:strong-identify}) are also well-fit by a combination of three Gaussian functions.

%= = = = = = = = = = = = = = = = = = = = = = = = = = = = = = = = = = = = = = 
\subsubsection{Combining Lensing Regimes}\label{sec:LF-combine}

All of the fields have a weak lensing magnification PDFs based on their overdensity (see Section~\ref{sec:weak}), but we have also identified one strongly-lensed candidate and three dropouts close to large foreground galaxies that produce an intermediate magnification PDF (see Section~\ref{sec:strong-identify}).

To account for the magnification bias, we need to use the correct magnification PDF for each field. In the case when a strong or intermediate lens appears present, we split the field into two parts for the calculation of the posterior: one is a circle with radius $10 \; \theta_\textrm{ER}$ containing the dropout and the deflector, where we use the strong or intermediate lens magnification PDF. For the remainder of the field we use the weak lensing magnification PDF.

Whilst the total flux across the sky is conserved, locally over- or underdensities that produce magnification not only magnify fluxes, but also increases areas. Hence, the individual BoRG fields we observe have been magnified (or demagnified) from their true sizes. We account for this in the posterior probability \Eq{eqn:LF-bayesian_bayes-margpost} by dividing the measured area of each field by the mean magnification in that field, $\overline{\mu}_l$ from the magnification PDFs. For weak lensing magnification PDFs $\langle\mu_l\rangle \sim 1$. For the intermediate lensing case $1.4 < \overline{\mu}_l < 2$ due to our selection process.

As magnification is most important for the bright-end of the
LF, and negligible at the faint end, for simplicity
and without loss of precision, we adopt $\mu=1$ for the 59 fainter
dropouts \citep{Bouwens2011}.  Additionally, one of the BoRG fields
(borg\_1815\_3244) is centered on the Galactic plane and is dominated
by stars. We discard this field in our calculation of the LF. 
%\NB{I thought we were discarding this one, no?}

% ===========================================================================
\section{Results}\label{sec:results}

Using the framework described in Section~\ref{sec:LF} to account for the magnification bias we present our estimation of the $z\sim8$ galaxy LF based on our sample of 97 $z\sim8$ LBGs (described in Section~\ref{sec:data}).
First, in Section~\ref{sec:occur} we compare our estimates of strong and intermediate lensing probabilities with the actual observations. Then, in Section~\ref{sec:results-LF} we carry out the inference of the $z\sim8$ LF. Finally, in Section~\ref{sec:results-LF-z}, we use our semi-analytical model of strong lensing optical depths described in Section~\ref{sec:strong} to predict the form of observed LFs at $z \geq 8$.

%= = = = = = = = = = = = = = = = = = = = = = = = = = = = = = = = = = = = = = 
\subsection{Strong and Intermediate Lensing Events in the BoRG Survey}\label{sec:results-borg-strong}
\label{sec:occur}

The simple SIS strong lensing model described in Section~\ref{sec:strong-zevol} predicts the probability of $z\sim8$ sources in the BoRG survey being multiply imaged to be $\sim 3-15\%$, increasing as the field limiting magnitude becomes brighter than $M^\star$. The majority of the BoRG fields have a multiple-image probability for high-redshift sources of $< 10\%$ (see Figure~\ref{fig:strong-zevol_flens}). We predict that 1-2 of the 38 BoRG Y-band dropouts may be strongly lensed.

One candidate strong lens system in BoRG was presented by \citet{Barone-Nugent2013a}, a rigorous search for strong lenses in all 71 BoRG fields as part of this work revealed one more candidate. Additionally, this search revealed three candidate intermediate lens systems, with $\mu>1.4$. These candidates are presented in Figure~\ref{fig:strong-postage} and Table~\ref{tab:strong-strongish}. Whilst strong lensing creates larger magnification, the probability of encountering a strong lens along the line-of-sight is low: as shown in Figure~\ref{fig:strong-zevol_optdep-diff-vdfevol} the optical depth is roughly $\tau\approx 0.31\%$ for a source at $z=8$. The optical depth for intermediate lensing is much higher: for an object to experience intermediate lensing it must be within $3.5\theta_\textrm{ER}$ of the foreground deflector, resulting in $\tau\approx 4\%$ for a source at $z=8$. Thus, intermediate lensing offers an additional boost to the flux of high-redshift galaxies, and must be correctly accounted for in estimations of the LF.

%= = = = = = = = = = = = = = = = = = = = = = = = = = = = = = = = = = = = = = 
\subsection{Inference of the Intrinsic $z\sim8$ LF}\label{sec:results-LF}

We estimate the $z\sim8$ LF from the sample of 97 LBG described in Section~\ref{sec:data}, including the 38 S/N$_\textrm{J} > 5$ objects from the BoRG survey, including the effects of magnification bias. We sample the posterior distribution function for the Schechter function parameters with an MCMC chain of 40 000 steps.

\begin{figure}
\includegraphics[width=0.49\textwidth]{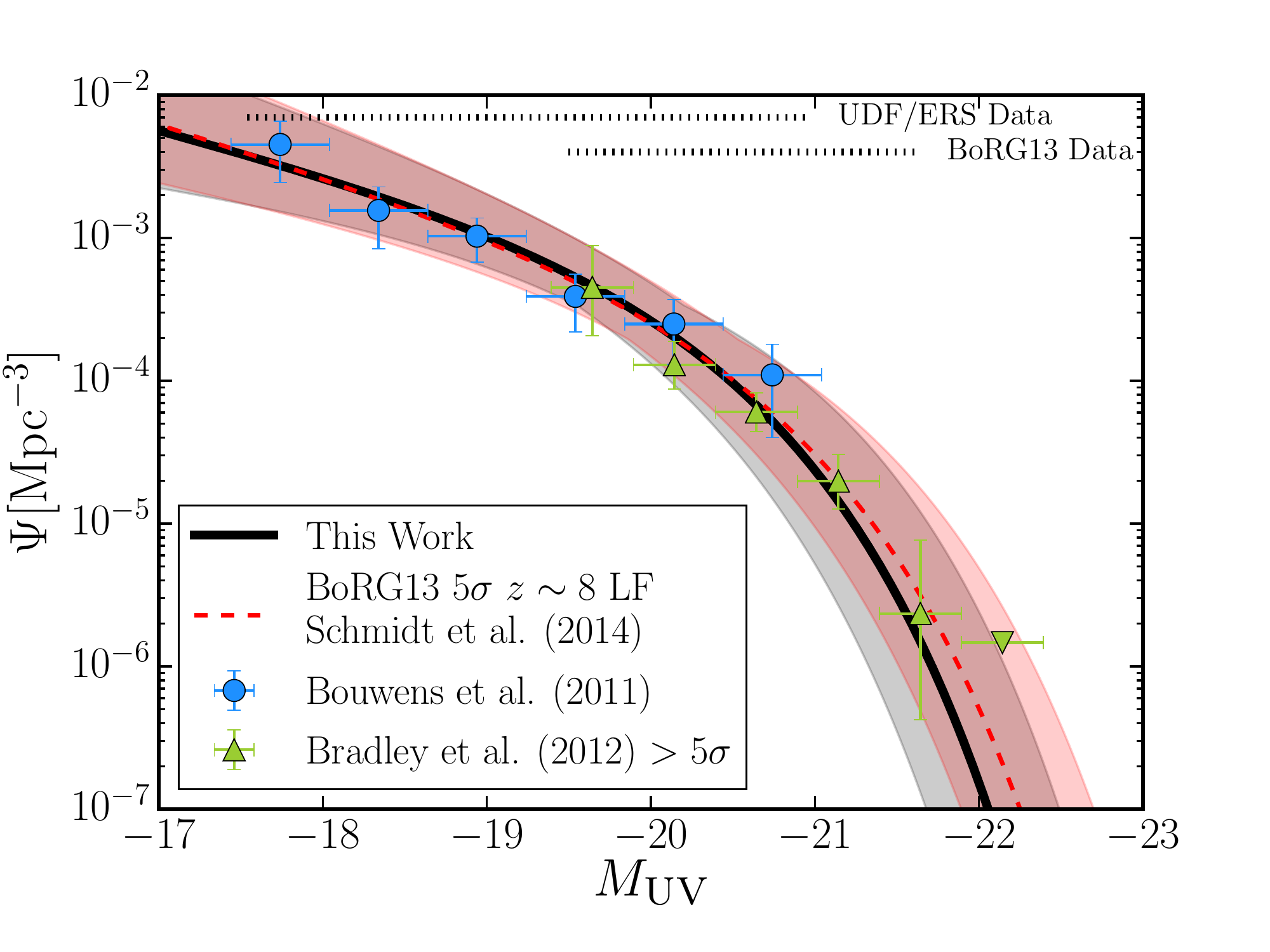}
\caption{The intrinsic $z\sim8$ LF, which is well-described by a \citet{Schechter1976} function, including the magnification bias due to weak and intermediate lensing in all BoRG fields (solid black line). We plot the LF without the treatment of the magnification bias \citep{Schmidt2014} for comparison (dashed red line). The lines corresponds to the median values of the MCMC samples and the shaded regions correspond to the 68\% confidence region of the samples. The LF estimated here is virtually indistinguishable from that of \citet{Schmidt2014}, demonstrating that magnification bias is not a significant effect at $z\sim8$. The Schechter parameters for this LF are given in Table~\ref{tab:results} along with literature values. The binned data from BoRG12 \citep{Bradley2012} and the faint HUDF/ERS candidates \citep{Bouwens2011} are also plotted as blue and green points respectively. The inverted green triangle denotes the brightest BoRG dropout. We note that the LF is estimated from the unbinned data.}
\label{fig:results_LF-borg}
\end{figure}

\begin{figure*}
\includegraphics[width=0.98\textwidth]{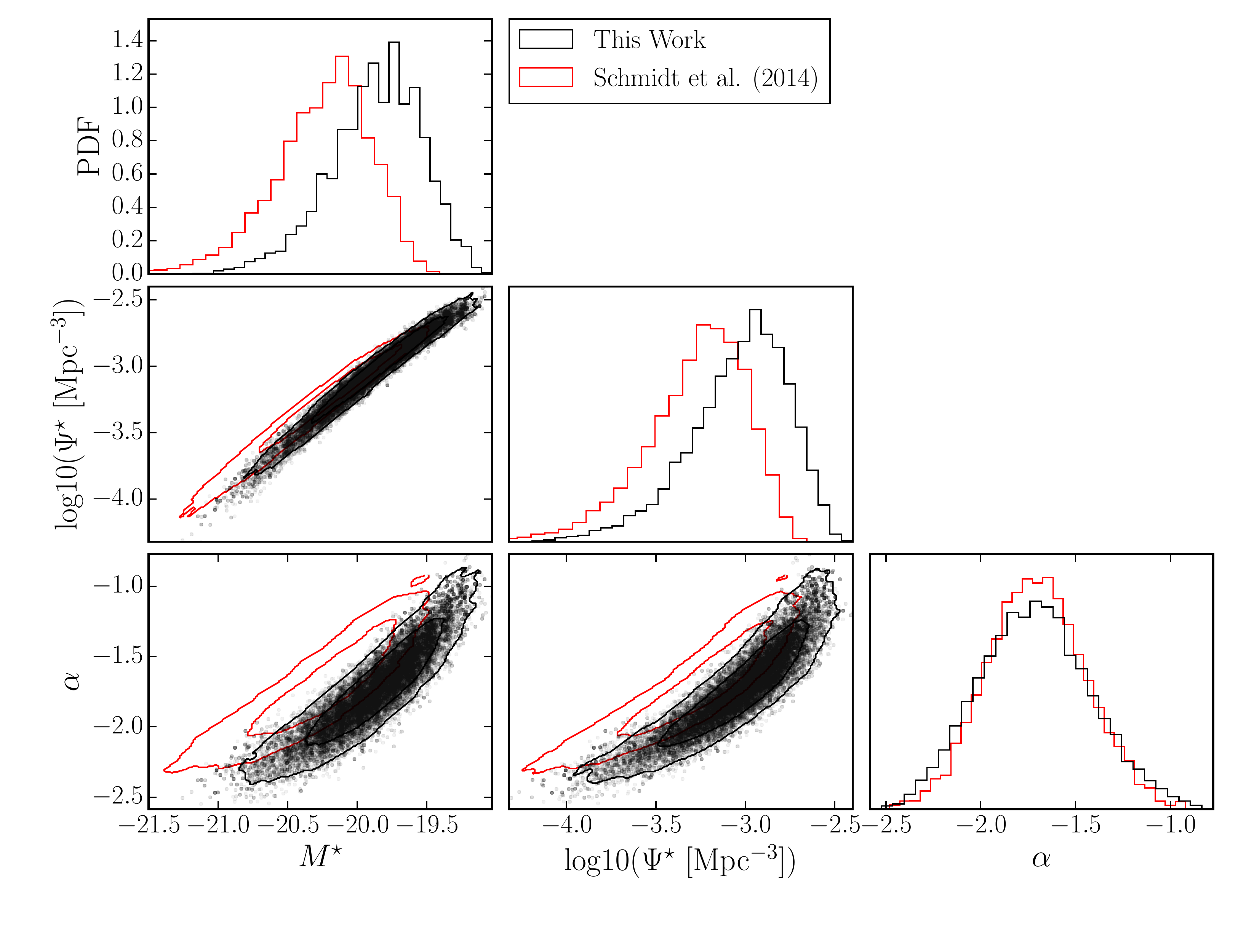}
\caption{The correlations between the $z\sim8$ LF Schechter function parameters $(\alpha, \, M^\star \textrm{ and } \Psi^\star$) estimated from the BoRG dropouts including treatment of magnification bias (black), compared to the parameters obtained without the treatment of magnification bias \citep[red,][]{Schmidt2014} with $1\sigma$ and $2\sigma$ confidence contours. There is clear correlation between all three parameters. The top panels show the marginalized PDFs for each parameter.}
\label{fig:results_LF-borg-params}
\end{figure*}

The results of the estimated LF are shown in Figure~\ref{fig:results_LF-borg}, and the correlations between the Schechter function parameters and their PDFs are shown in Figure~\ref{fig:results_LF-borg-params}. We plot the results of \citet{Schmidt2014} for comparison in both figures. We see a small deviation from the uncorrected LF of $\sim0.15$ mag at the limit of the brightest BoRG source, and there is negligible difference between the LFs at $M > -21$. The Schechter function parameters for the new LF are within the uncertainties of the estimation by \citet{Schmidt2014}, though we find a slightly fainter value of $M^\star$ and higher value of $\Psi^\star$ than \citet{Schmidt2014}. This is expected because of the slight deviation at the bright end of the LF, and there is a strong correlation between these parameters, as shown in Figure~\ref{fig:results_LF-borg-params}. It is clear that magnification bias is not a significant effect at this redshift and the luminosity range of the BoRG sources. This also demonstrates that although we predict $3-15\%$ of the BoRG sources are strongly lensed this does not affect the LF within the survey limits, as predicted in Section~\ref{sec:theory-magbias}.

Our results are in good agreement with those of Fialkov \& Loeb (2015) who use an independent semi-analytic method to show the effect of magnification bias is small below $M>-21.5$. Fialkov \& Loeb (2015) predict that if the brightest observed galaxy has absolute magnitude $M_\textsc{uv} = - 24.5$ - lying in the significantly distorted tail of the magnified LF (Wyithe et al. 2011) - there is a $\sim13.3\%$ discrepancy in the normalization of a Schechter LF at $z\sim8$ for sampled galaxies with $\mu_\textrm{max} = 2$ (i.e. only weak and intermediate lensing effects) compared to the intrinsic LF. Whilst this upper limit is several orders of magnitude brighter than currently observed, this demonstrates that it will be important to include the effects of magnification bias from weak and intermediate lensing in surveys that find extremely bright galaxies.

Table~\ref{tab:results} summarizes the estimated Schechter function
parameters for this LF in comparison with other
recent LF estimates from the literature. We find that
our fit parameters are in good agreement with the recent literature,
demonstrating that magnification bias is not affecting current
$z\sim8$ LF observations. Note that our results have
significantly smaller error bars than those of \citet{Finkelstein2014},
because their sample contains only 3 $z\sim 8$ galaxies brighter than $M
= -21$, making their fit less well-constrained at the bright end.

Our results show that magnification bias does not affect current estimates of the LF 
at $z \simlt 8$ and therefore cannot explain the apparent flattening of the bright-end 
of the LF recently observed by \citet{Bowler2014b,Bowler2014a} 
and \citet{Finkelstein2014} at $z\sim 7 - 8$. \citet{Bowler2014b,Bowler2014a}
accounted for strong lensing of bright sources, but they still find a deviation of $\sim0.4$ mag from a Schechter fit at $M=-22$. We predict a lensed fraction of $\sim 3-15 \%$ for 
bright galaxies (Figure~\ref{fig:strong-zevol_flens}) from the BoRG survey which is essentially 
free of cosmic variance \citep{Trenti2008}, so providing cosmic variance and contamination by lower redshift interlopers \citep{Hayes2012,Bradley2012,Bouwens2013,Schmidt2014}
were correctly accounted for in the work of \citet{Bowler2014b,Bowler2014a} 
and \citet{Finkelstein2014}, we expect the magnification bias to be 
negligible in the bright-end of these LFs. This lends credence to the interpretation that these observations may be 
the result of the changing intrinsic properties of galaxies at $z\simgt 7$, possibly due to 
changing dust fractions \citep{Cai2014} and/or feedback processes \citep{Somerville2012}.

% ---------------------------------------------------------------------------
% ---------------------------------------------------------------------------

% ---------------------------------------------------------------------------
% ---------------------------------------------------------------------------

%= = = = = = = = = = = = = = = = = = = = = = = = = = = = = = = = = = = = = = = = 
\begin{table*}
\centering{
\caption[ ]{Comparison of $z\sim8$ Schechter LF Parameters}
\label{tab:results}
\begin{tabular}[c]{lllll}
\hline
\hline
Reference &  $M^\star$ & $\alpha$ & $\log_{10} \Psi^\star$ [Mpc$^{-3}$] \\
\hline
This work 						& $-19.85^{+0.30}_{-0.35}$ 	&  $-1.72^{+0.30}_{-0.29}$  &  $-3.00^{+0.23}_{-0.31}$  \\
\citet{Finkelstein2014}			& $-20.89^{+0.74}_{-1.08}$	& $-2.36^{+0.54}_{-0.40}$	& $-4.14^{+0.65}_{-1.01}$	\\
\citet{Bouwens2014}				& $-20.63\pm0.36$			& $-2.02\pm0.23$ 			& $-3.68\pm0.32$			\\
\citet{Schmidt2014} $5\sigma$ 	& $-20.15^{+0.29}_{-0.38}$ 	& $-1.87^{+0.26}_{-0.26}$ 	& $-3.24^{+0.25}_{-0.34}$  \\
\citet{Schmidt2014}  $8\sigma$ 	& $-20.40^{+0.39}_{-0.55}$	& $-2.08^{+0.30}_{-0.29}$  	& $-3.51^{+0.36}_{-0.52}$  \\
\cite{McLure2013}      			& $-20.12^{+0.37}_{-0.48}$ 	& $-2.02^{+0.22}_{-0.23}$ 	& $-3.35^{+0.28}_{-0.47} $ \\
\citet{Schenker2013}			& $-20.44^{+0.47}_{-0.35}$ 	& $-1.94^{+0.21}_{-0.24}$ 	& $-3.50^{+0.35}_{-0.32} $ \\
\hline
\end{tabular}}
\end{table*}
%= = = = = = = = = = = = = = = = = = = = = = = = = = = = = = = = = = = = = = = = 

%= = = = = = = = = = = = = = = = = = = = = = = = = = = = = = = = = = = = = = 
\subsection{Predictions for $z>8$ and Future Surveys}\label{sec:results-LF-z}

There is clear evolution in the LF for $z<8$ \citep[e.g.][]{Bouwens2007,VanderBurg2010,Bouwens2014,Bowler2014b,Finkelstein2014}, and this is expected to continue to higher redshifts. However, the processes which drive this evolution are not well-understood: the evolution is thought to follow hierarchical structure formation and the evolution of the halo mass function \citep{Vale2004}, but there are also important quenching processes that may reduce star formation in massive galaxies \citep{SaasFee,Somerville2012}, and changes in the amount of dust present in galaxies will affect the attenuation of flux. Thus there are a multitude of theoretical models for the evolution of the LF.

The gravitationally lensed LF (\Eq{eqn:theory-magbias_mod-LF}) exhibits a significant `kick' in the bright-end tail for $M \simlt -22$ at $z\sim8$. This is just beyond the brightest BoRG objects, so it is unlikely that the BoRG survey observes the regime of magnification bias at the bright-end. This is in agreement with theoretical studies by \citet{Wyithe2011} and \citet{Fialkov2015}. However, in upcoming wide-area surveys magnification bias presents a useful tool to test LF evolution models because it allows us to probe the bright end, where there are large theoretical uncertainties and the evolution is expected to be fast \citep{Bowler2014b}.

\begin{figure*}[!h]
\centering{
\includegraphics[width=0.49\textwidth]{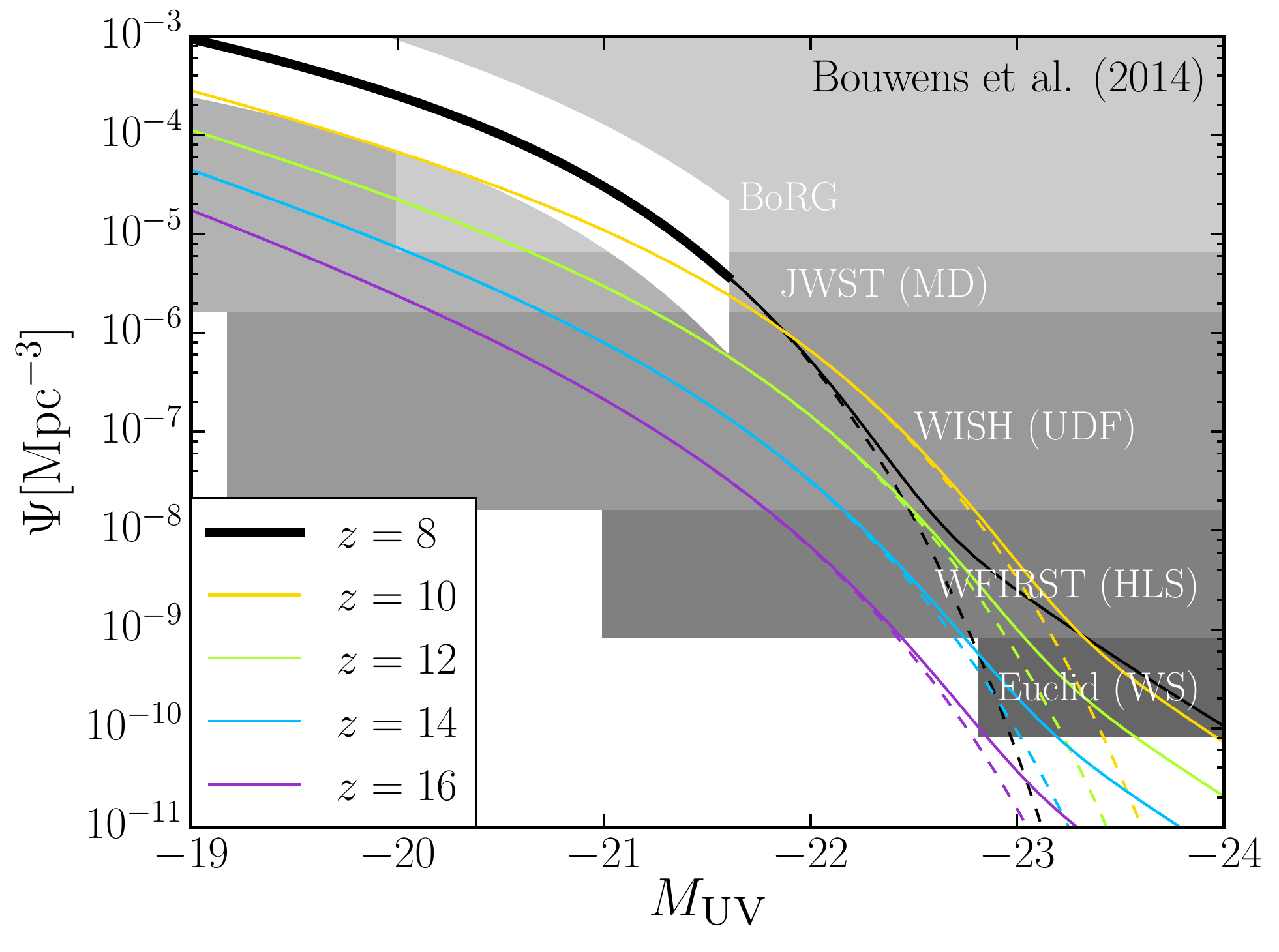}
\includegraphics[width=0.49\textwidth]{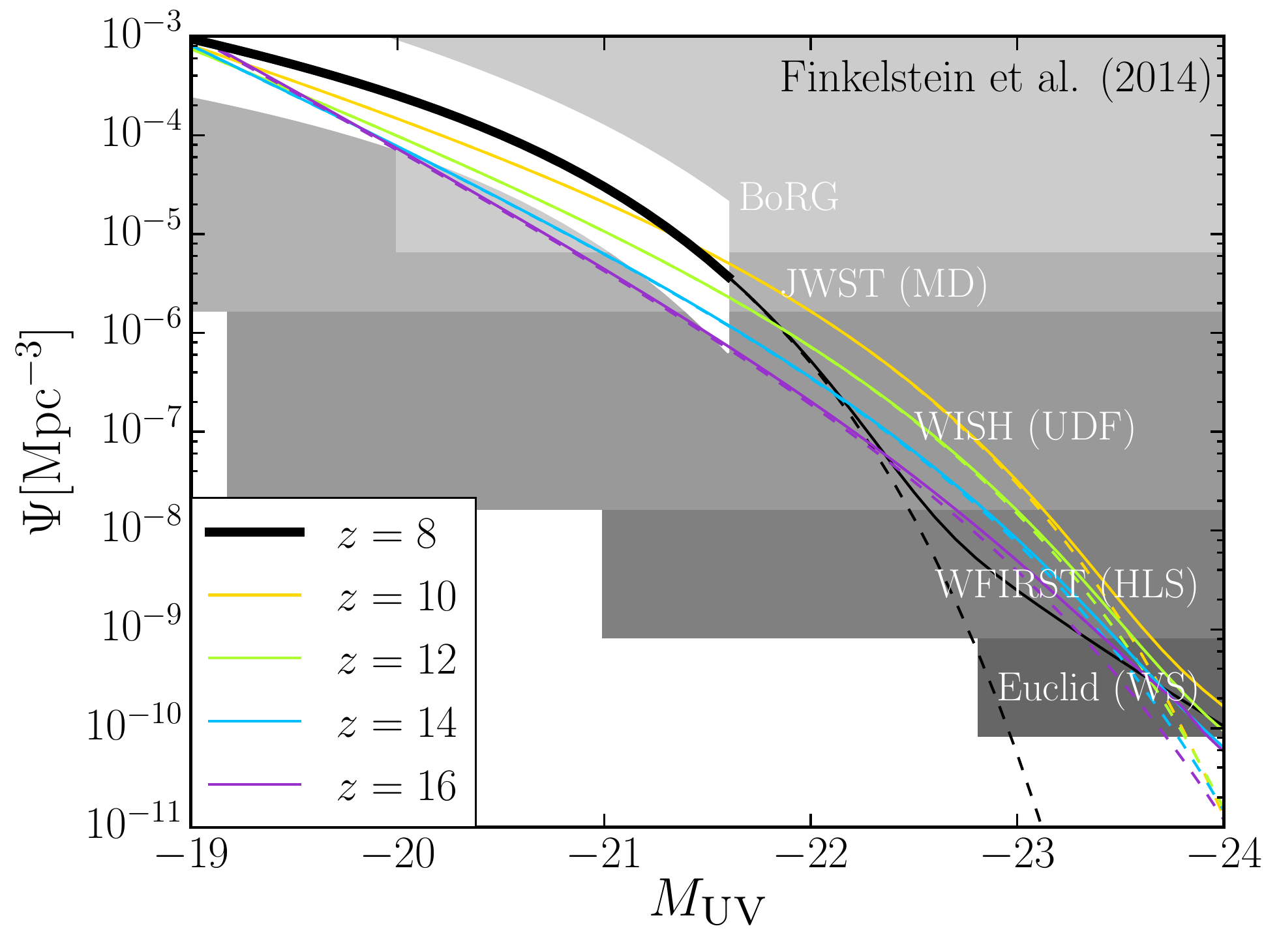} 
\includegraphics[width=0.49\textwidth]{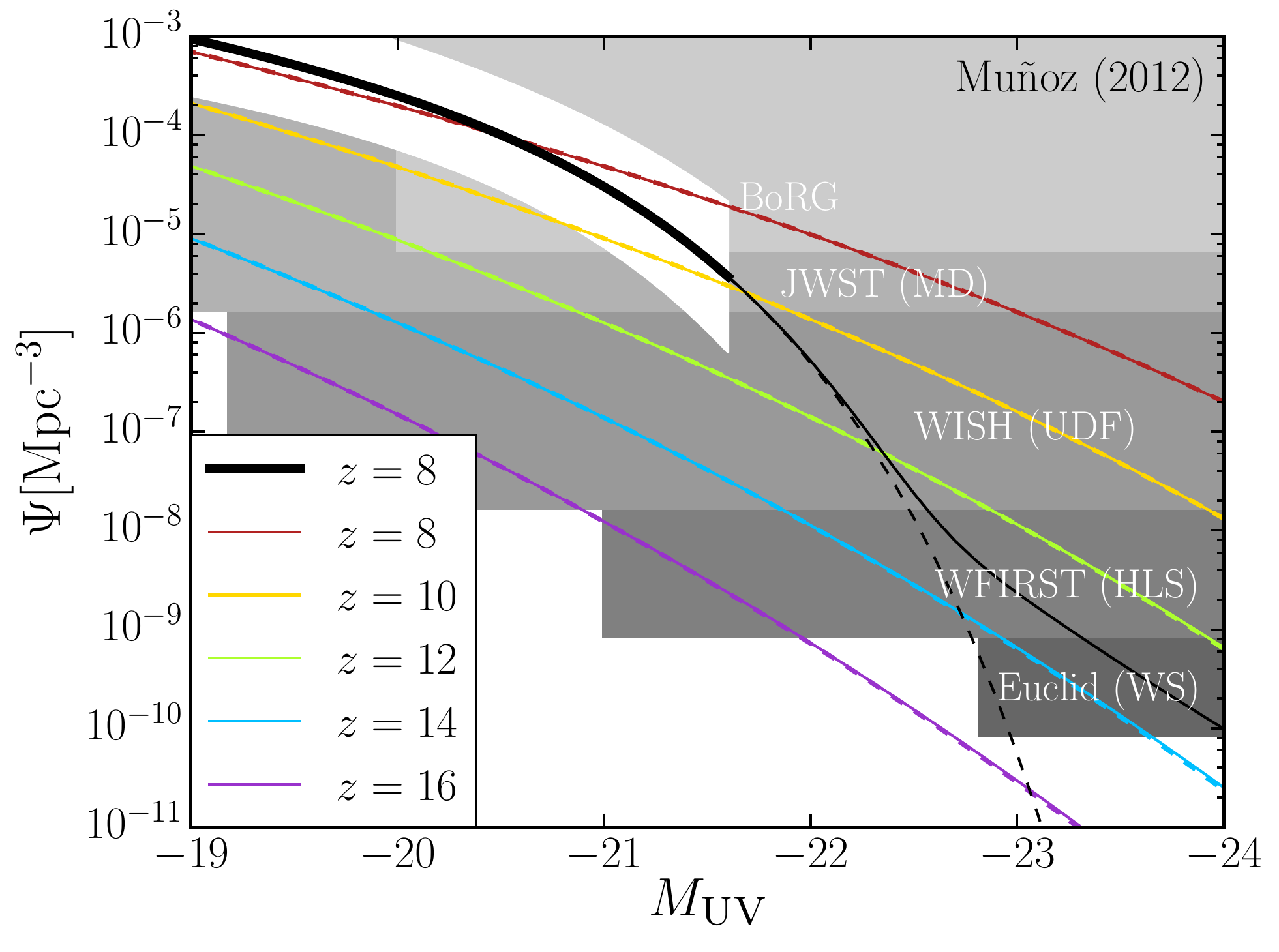}
\includegraphics[width=0.49\textwidth]{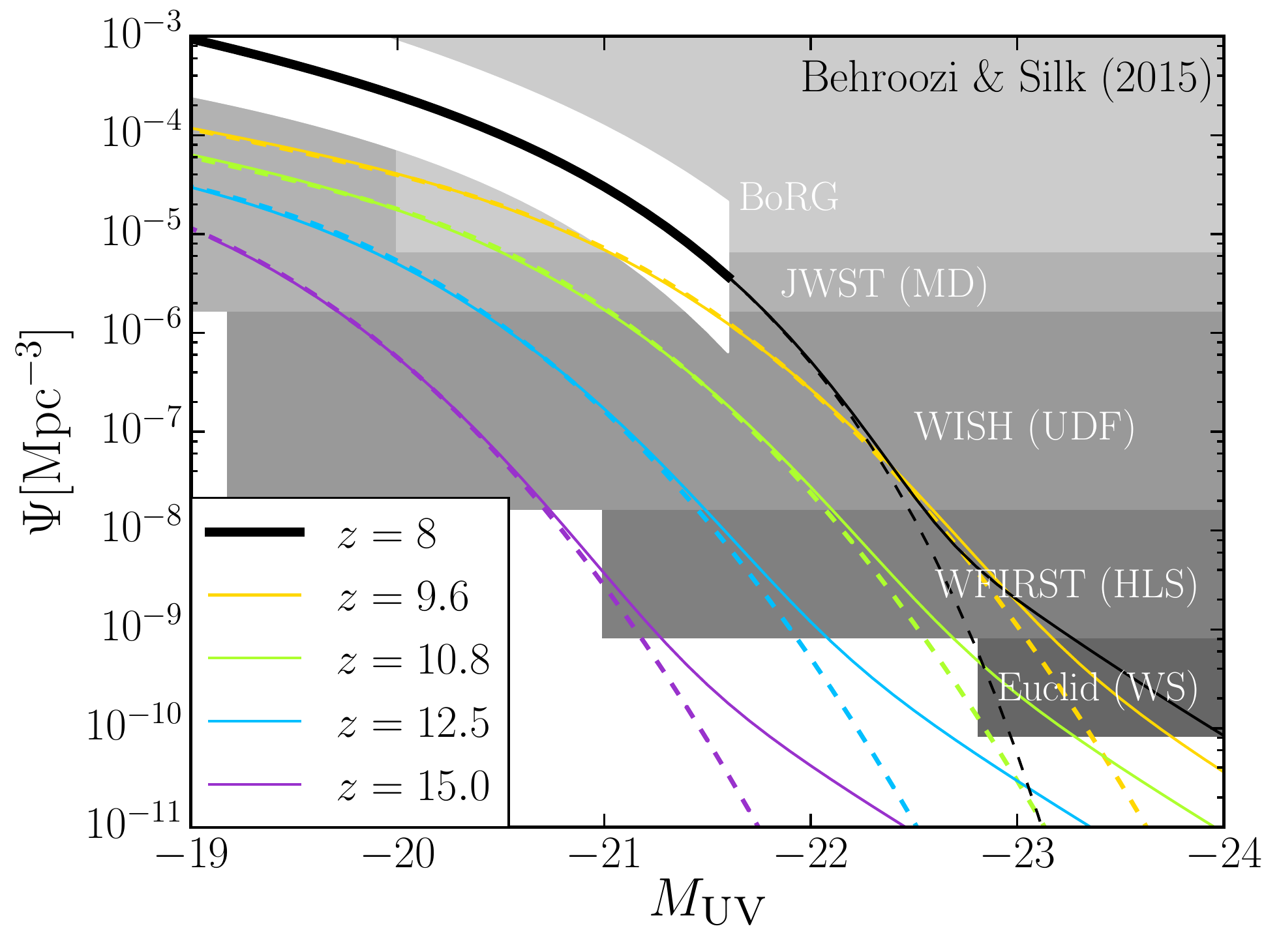}
}
\caption{Predicted observed LFs for $z\geq 8$ redshifts. For $z=8$ we use the Schechter LF from \citet{Schmidt2014}, plotted as a thick black line. The white band indicates the error on the Schechter function parameters, and the thin black line is the extrapolation of the LF beyond the observational limit. We show the regions of magnitude and volume observable by current and future surveys: the total BoRG survey including the $z\sim 8$ survey described in Section~\ref{sec:data-borg} and the upcoming BoRG $z\sim 9,10$ survey (HST Cycle 22, PI: Trenti); the James Webb Telescope Medium Deep (JWST MD) \citep{Windhorst2006}; the Wide-Field Imaging Surveyor for High-Redshift Ultra-Deep Field (WISH UDF, http://wishmission.org/en/doc.html); the Wide-Field Infrared Survey Telescope High Latitude Survey (WFIRST HLS) \citep{Spergel2013} and the Euclid Wide Survey (WS) \citep{Laureijs2011}.  As explained in the text, BoRG does not survey enough area to observe the rarest bright sources which are most affected by magnification bias, but future wide-field surveys will be dominated by this effect. {\bf (Top Left)} For $z>8$ we use the LF model from \citet{Bouwens2014} which is an extrapolation from $z\sim10$. {\bf (Top Right)} For $z>8$ we use extrapolate the evolution of the Schechter function parameters over $4<z<8$ from \citet{Finkelstein2014}. {\bf (Bottom Left)} For $z\geq 8$ we use the luminosity model from \citet{Munoz2012} which is based on the evolution of the halo mass function. These do not exhibit the sharp cut-off at the bright-end and are not affected by magnification bias. {\bf (Bottom Right)} For $z>8$ we use the LF evolution model from \citet{Behroozi2015}. The dashed lines indicate the intrinsic LFs, the solid lines are the observed LFs including the magnification bias calculated using \Eq{eqn:theory-magbias_mod-LF}.}
\label{fig:results_LF-zevol}
\end{figure*}

In order to explore the range of possible scenarios, in Figure~\ref{fig:results_LF-zevol} we plot the predicted intrinsic (dashed lines) and observed (solid lines) LFs for a range of redshifts, comparing a variety of evolution models. We assume these models are the intrinsic LFs at a given redshift and used \Eq{eqn:theory-magbias_mod-LF} to estimate the observed LF. We plot the BoRG $z\sim8$ LF \citep{Schmidt2014} for comparison. Additionally, we mark the comoving volumes and magnitude ranges accessible to future high-redshift surveys.

The top left panel shows the LF model from
\citet{Bouwens2014} which is an extrapolation from observations at $z <
10$. The top right panel shows the LF model from
\citet{Finkelstein2014} which is an extrapolation from observations at $4 < z <
8$. The bottom left panel shows the model developed by \citet{Munoz2012}
which follows the evolution of the halo mass function, and includes dust attenuation.  The bottom right panel is a model from \citet{Behroozi2015} constructed from a comparison of the specific
star formation rate to the specific halo mass accretion rate, and
including dust models from \citet{Charlot2000}. The four models have
significantly different behaviors at the bright end. While the
\citet{Bouwens2014} model has by construction a bright end that is very
similar to that measured at lower redshifts, the \citet{Munoz2012}
model has a very shallow bright end, and the \citet{Finkelstein2014} and \citet{Behroozi2015} models are in-between. As a result, the effects of magnification bias (which are
stronger for the steeper LF) are very different: negligible in the
\citet{Munoz2012} case and appreciable in the three other cases.
However, the bright end of the \citet{Munoz2012} model is the easier
one to test observationally, within reach of a James Webb Space
Telescope medium depth, medium width survey \citep[e.g. JWST MD][]{Windhorst2006}.

Except in the case of a very shallow bright end, we do not expect the magnification bias 
to be significant in our upcoming BoRG $z\sim 9,10$ survey (HST Cycle 22, PI Trenti). 
In all cases, it is clear that surveys covering $>100$ deg$^2$,
e.g. Euclid and WFIRST, should find many bright $z > 8$ LBGs. We
expect the observed high-redshift galaxy samples will be dominated by
magnification bias in these surveys. We predict almost all $z\sim8$
sources in Euclid will have been strongly lensed. The framework
developed in this work will be crucial for determining the intrinsic
luminosity of high-redshift sources found in such surveys.

Our results confirm the suggestion by \citet{Wyithe2011} that
magnification bias will be important to probe the bright end of the LF
at high redshift. However, we find that the magnitude of the effect is
less pronounced than in that study, owing mostly to our accounting for
the redshift evolution of the deflector population.

% ---------------------------------------------------------------------------
% ---------------------------------------------------------------------------

% ====================================================================== 
\section{Summary and Conclusion}\label{sec:conc}

We have introduced a systematic way to account for the magnification bias in estimations of high-redshift LFs. The method involves estimating the probability density function for weak lensing magnification along a given line-of-sight by comparison with results from the reconstruction of simulated halo data, and by estimating the strong and intermediate lensing magnification PDF of dropouts due to massive deflector galaxies in close proximity to the dropout.

We applied this method to estimate the $z\sim8$ LF from the 38 BoRG Y-band dropouts and 59 fainter dropouts from \citet{Bouwens2011}. Our main results are summarized as follows:

\begin{enumerate}[(a)]
   \item The probability of a BoRG $z\sim8$ dropout being multiply imaged is $\sim 3-15$\%, increasing with limiting magnitude. This is consistent with finding two strongly-lensed dropouts in the BORG survey: the candidate system presented in \citet{Barone-Nugent2013a}, and the additional strongly-lensed candidate dropout in this paper. We also find three dropouts which may experience significant magnification without multiple imaging, consistent with our expectations.
   \item We extended the Bayesian formalism for the estimation of the LF parameters presented by \citet{Schmidt2014} to account for the magnification bias. This involves marginalizing over the magnification PDFs for strong and weak lensing effects. The inferred Schechter function parameters are:
      \begin{itemize}   
      \item[] $M^\star = -19.85^{+0.30}_{-0.35}$,
      \item[] $\alpha = -1.72^{+0.30}_{-0.29}$, 
      \item[] $\log_{10} \Psi^\star [\textrm{Mpc}^{-3}] = -3.00^{+0.23}_{-0.31}$,
      \end{itemize}
      These values do not differ significantly from estimates not accounting for the magnification bias.
   \item Thus magnification bias cannot be an explanation for the apparent flattening of the bright-end of the LF recently observed by \citet{Bowler2014b,Bowler2014a} and \citet{Finkelstein2014}.
   \item The $z\sim8$ LF appears significantly magnified for extremely bright galaxies ($M_\textsc{uv} < -22$). Though current surveys have not observed such rare, luminous galaxies, future wide-field surveys will probe this region. For surveys $> 100$ deg$^2$, e.g. WFIRST, Euclid, we predict that samples of $z\gtrsim8$ galaxies will be dominated by magnification bias.
   \item Magnification bias will be a useful tool to distinguish between high-redshift LF evolution models. In particular it could help determine whether the LF transitions from a Schechter form to a power-law form at high redshift, indicating significant changes in the astrophysical properties of those galaxies.

\end{enumerate}

% ====================================================================== 
\acknowledgments

We thank Joey Mu\~{n}oz for useful discussions and providing his LF evolution model; Peter Behroozi for providing his LF evolution model; Sirio Belli for providing photometry of the galaxies described in \citet{Belli2014,Belli2014a}; and Stefan Hilbert for useful comments regarding the weak lensing simulations.

This work was supported by the HST BoRG grants GO-11700, 12572, and 12905. This paper is based on observations made with the NASA/ESA Hubble Space Telescope, obtained at the Space Telescope Science Institute.

This work made use of the freely available Pangloss code, written by Tom Collett and Phil Marshall. The Millennium Simulation databases used in this paper are publicly available through the German Astrophysical Virtual Observatory.

% ====================================================================== 

\begin{appendix}

% = = = = = = = = = = = = = = = = = = = = = = = = = = = = = = = = = = = = = = 
\section{Bayesian Framework for Estimating the luminosity function}\label{app:bayes}
We use Bayesian statistics to find the relationship between the prior probability of the $z \sim 8$ dropouts being galaxies with LF Schechter parameters $\theta = (\alpha, \;  L^\star, \;  \Psi^\star)$, and these parameters' posterior probability given the dropout candidates' detection threshold in the J-band, assuming their non-detection in the V-band. The posterior probability is given by:

   \BE  \label{eqn:bayes}
      p(\theta \,| \,L_\textrm{J,obs}, I_V =0) \propto p(\theta) \times p(L_\textrm{J,obs}, I_V = 0 \,|\, \theta)
   \EE
where the last term is the likelihood and $p(\theta)$ is the prior on the LF parameters. We will assume uniform priors on $\alpha$ and $\log_{10} L^\star$.

We can expand the expression for the posterior:
   \BEA \label{eqn:bayes-post}
      p(\theta \,|\, L_\textrm{J,obs},I_\textrm{V}=0) \propto \; p(\theta)&&\; C^{N_z}_{n_z} \;  \prod_{l}^\mathcal{C}  \left[1- A_l/A_\textrm{sky}\; p(I=1|\theta) \right]^{N_z-c_{lz}} \times \; \prod_i^{n_z} p(L_{\textrm{J,obs},i}\,|\,\theta) \nonumber \\     
      \times&& \;  C^{N_c}_{n_c} \; \prod_{l}^\mathcal{C} \left[1- A_l/A_\textrm{sky}\; p(I=1|\theta) \right]^{N_c-c_{lc}} \times \; \prod_i^{n_c} p(L_{\textrm{J,obs},i}\,|\,\theta)
   \EEA
where the $C^a_b$ terms are binomial coefficients which correctly model the distribution of source counts. $N_z$ and $N_c$ are the number of high-redshift sources given the intrinsic LF and the number of potential contaminants in the Universe respectively. We will assume a uniform prior on $\log_{10} N_z$. In the observed sample the number of high-redshift sources and contaminants are given by $n_z$ and $n_c$. The total number of galaxies in the observed sample, $n_t$ is given by their sum. We take the product over $\mathcal{C}$ individual observed fields where $c_l$ represents the number of galaxies in the $l$'th field with $n_t$ also given by the sum of $c_l$ over all of the fields. The fraction of the sky covered by the $l$'th field is given by $A_l/A_\textrm{sky}$. The contamination fraction in each field, $f_l$ is set at the fiducial value of 42\% \citep{Schmidt2014,Bradley2012} for the BoRG sources, the contamination fraction for the fainter HUDF/ERS sources \citep{Bouwens2011} is included in the selection function (see below).

The last term in \Eq{eqn:bayes-post} is the likelihood for the $i$'th object in the sample. In \citet{Schmidt2014} this was expressed as:
   \BEA   \label{eqn:bayes-likelihood-kasper}
      p(L_\textrm{J,obs}\,|\,\theta) &=& \; \int_0^\infty p(L_\textrm{J,obs} \,|\, L_\textrm{J,true}) \; p(L_\textrm{J,true} \,|\, \theta )\; dL_\textrm{J,true} \\
      &=& \;  \int_0^\infty \mathcal{N}(L_\textrm{J,obs} \,|\, L_\textrm{J,true}, \, \delta L_\textrm{J,field})\; \textrm{gamma}(L_\textrm{J,true} \,|\, \alpha,L^\star) \; dL_\textrm{J,true} \; \nonumber   
   \EEA
where we use $p(L\,|\,\theta)\propto \frac{\Psi(L, \theta)}{\Psi^{\star}}$ (see Equation~(1) of \citet{Kelly2008}). The function gamma$(L_\textrm{J,true} \,|\, \alpha,L^\star)$ is related to the Schechter LF (\Eq{eqn:theory_LF-sch}) as $\textrm{gamma}(L \,|\, \alpha,L^\star) = \frac{\Psi(L)}{\Psi^\star\Gamma(\alpha+1)}$.

   \BE  \label{eqn:bayes-errgauss}
      \mathcal{N'}(L_\textrm{J,obs} \,|\, L_\textrm{J,true}, \, \delta L_\textrm{J,field}) =  \frac{1}{\delta L_\textrm{J,field}\sqrt{2\pi}} \exp\left[-\frac{(L_\textrm{J,obs}-L_\textrm{J,true})^2}{2\, \delta L_\textrm{J,field}^2} \right]
      \EE
represents the true luminosity inferred from the observations assuming a Gaussian measurement error with $\delta L_\textrm{J,field}$ being the median photometric error in the J-band in the given field.

In order to include the effects of the magnification bias, we must integrate over the nuisance parameter $L_\textrm{J,mag}$, which represents the luminosity of an object in the J-band, magnified above its true luminosity. Including this \Eq{eqn:bayes-likelihood-kasper} becomes:

   \BE  \label{eqn:bayes-Lobs-like}
      p(L_\textrm{J,obs}\,|\,\theta) = \int \int p(L_\textrm{J,obs}\,|\,L_\textrm{J,mag})\; p(L_\textrm{J,mag}\,|\,L_\textrm{J,true})\; p(L_\textrm{J,true}\,|\,\theta) \; dL_\textrm{J,true} \; dL_\textrm{J,mag}
   \EE
where $p(L_\textrm{J,obs}\,|\,L_\textrm{J,mag})$ is now the term with Gaussian measurement errors similar to \Eq{eqn:bayes-errgauss}, given that we make observations of magnified luminosities:
   \BE  \label{eqn:bayes-errgauss2}
      \mathcal{N}(L_\textrm{J,obs} \,|\, L_\textrm{J,mag}, \, \delta L_\textrm{J,field}) =  \frac{1}{\delta L_\textrm{J,field}\sqrt{2\pi}} \exp\left[-\frac{(L_\textrm{J,obs}-L_\textrm{J,mag})^2}{2\, \delta L_\textrm{J,field}^2} \right]
   \EE

To find the probability that luminosity is magnified from its true luminosity, $p(L_\textrm{J,mag}\,|\,L_\textrm{J,true})$, we must integrate over the full magnification probability density:
   \BEA  \label{eqn:bayes-LmagLtrue}
         p(L_\textrm{J,mag}\,|\,L_\textrm{J,true}) &=& \int p(L_\textrm{J,mag} \,|\, \mu, L_\textrm{J,true} )\; p(\mu) \; d\mu 
   \EEA

We can marginalize over $L_\textrm{J,mag}$ in the first part of \Eq{eqn:bayes-Lobs-like}:
   \BEA  \label{eqn:bayes-f}
   p(L_\textrm{J,obs} \,|\, L_\textrm{J,true}) &=& \int p(L_\textrm{J,obs}\,|\,L_\textrm{J,mag}) \; p(L_\textrm{J,mag}\,|\,L_\textrm{J,true})\;  dL_\textrm{J,mag}  \nonumber \\ 
         &=& \int \int p(L_\textrm{J,mag}\,|\,\mu, L_\textrm{J,true} )\; p(\mu)\; p(L_\textrm{J,obs},L_\textrm{J,mag},\,\delta L_\textrm{J,field})\;  d\mu \;  dL_\textrm{J,mag} \nonumber \\ 
         &=& \int \int \delta(L_\textrm{J,mag} - \mu L_\textrm{J,true} )\; p(\mu) \; \mathcal{N}(L_\textrm{J,obs}\,|\,L_\textrm{J,mag},\,\delta L_\textrm{J,field})\;  d\mu \;  dL_\textrm{J,mag} \nonumber \\ 
         &=& \int p(\mu) \; \mathcal{N}(L_\textrm{J,obs}\,|\,\mu L_\textrm{J,true}, \, \delta L_\textrm{J,field}) \;  d\mu \nonumber  \\         
         &=& \int p(\mu) \frac{1}{\delta L_\textrm{J,field}\sqrt{2\pi}} \exp \left[-\frac{\left(L_\textrm{J,obs} - \mu L_\textrm{J,true} \right)^2 }{2\, \delta L^2_\textrm{J,field}} \right] \; d\mu
   \EEA

Here we have used the Dirac delta function $\delta(L_\textrm{J,mag} - \mu L_\textrm{J,true} )$ to map true luminosities to magnified luminosities. To make computation of \Eq{eqn:bayes-post} feasible, we integrate \Eq{eqn:bayes-f} analytically and want to remove any $L_\textrm{J,mag}$ dependence we fit the magnification PDFs as a normalized linear combination of Gaussian terms with coefficients $\beta_i$ centered on $\overline{\mu}_\textrm{i,mag}$ with standard deviation $\sigma_\textrm{i,mag}$ :

   \BE  \label{eqn:bayes-pmu-Gauss}
      p(\mu) = \; \sum_i^n \beta_i \frac{1}{\sigma_\textrm{i,mag}\sqrt{2\pi}} \exp \left[-\frac{(\mu - \overline{\mu}_\textrm{i,mag})^2}{2\, \sigma^2_\textrm{i,mag}}\right]
   \EE    

\Eq{eqn:bayes-f} can then be integrated analytically: 
   \BE   \label{eqn:bayes-f-Gauss}
      p(L_\textrm{J,obs} \,|\, L_\textrm{J,true}) = \;  \sum_i^n \frac{\beta_i}{\sqrt{2\pi}} \frac{1}{\sqrt{\sigma^2_\textrm{i,mag} L^2_\textrm{J,true} + \delta L^2_\textrm{J,field}}} \exp \left[-\frac{(L_\textrm{J,obs} - \overline{\mu}_\textrm{i,mag} L_\textrm{J,true})^2}{2 \; (\sigma^2_\textrm{i,mag} L^2_\textrm{J,true} + \delta L^2_\textrm{J,field})} \right]
   \EE

As solid angle is also magnified in gravitational lensing we must divide the measured field area $A_l$ by the average magnification in each field $\overline{\mu}_\textrm{l}$. If $\overline{\mu}_\textrm{l} > 1$ the fields we observe appear larger than their true sizes.

We can therefore express \Eq{eqn:bayes-post} as (see \citet{Schmidt2014} for details):
   \BEA\label{eqn:bayes-margpost}
      p(\theta \,|\,  L_\textrm{J,obs},I_\textrm{V}=0) \propto \; &p&(\theta)\; \times \;   C^{N_z}_{(1-f)n_t} C^{\frac{f}{1-f} N_z}_{f n_t} \; \nonumber \\
      &\times&  \; \prod_{l}^\mathcal{C} \left[1- \frac{A_l}{\overline{\mu}_l A_\textrm{sky}}\; \int_0^\infty \int_0^\infty dL_\textrm{J,true,$l$}  \, dL_\textrm{J,obs,$l$} \; \mathcal{S}(L_\textrm{J,obs,l}) \; \mathcal{F}(L_\textrm{J,obs,$l$}, L_\textrm{J,true,$l$}) \right]^{\frac{1}{1-f_l}(N_z-(1-f_l)c_{l})} \nonumber \\
      &\times& \; \prod_i^{n_t} \int_0^\infty \mathcal{F}(L_\textrm{J,obs,$i$}, L_\textrm{J,true,$i$}) \;dL_\textrm{J,true,$i$}
   \EEA

Here we have defined $\mathcal{F}(L_\textrm{J,obs}, L_\textrm{J,true}) =  p(L_\textrm{J,obs} \,|\, L_\textrm{J,true})\; \textrm{gamma}(L_\textrm{J,true} \,|\, \alpha ,L^\star) $ and included the selection function $\mathcal{S}(L_\textrm{J,obs})$. The selection function estimates the completeness of the source selection and has been obtain for each individual BoRG field as explained in \citet{Oesch2012,Bradley2012} and \citet{Schmidt2014}.

Thus, \Eq{eqn:bayes-margpost} is the posterior probability distribution for a sample of $n_t$ binomially distributed objects, assumed have an intrinsic Schechter LF of the form shown in \Eq{eqn:theory_LF-sch}. The observed luminosity of each object is related to its true luminosity via a magnification PDF and an assumed Gaussian error distribution.

\end{appendix}

%======================================================================
\bibliographystyle{apj}
\bibliography{bibtexlibrary}

\begin{thebibliography}{}
\expandafter\ifx\csname natexlab\endcsname\relax\def\natexlab#1{#1}\fi

\bibitem[{{Atek} {et~al.}(2015){Atek}, {Richard}, {Kneib}, {Jauzac},
  {Schaerer}, {Clement}, {Limousin}, {Jullo}, {Natarajan}, {Egami}, \&
  {Ebeling}}]{Atek2015}
{Atek}, H., {Richard}, J., {Kneib}, J.-P., {et~al.} 2015, \apj, 800, 18

\bibitem[{Auger {et~al.}(2010)Auger, Treu, Bolton, Gavazzi, Koopmans, Marshall,
  Moustakas, \& Burles}]{Auger2010}
Auger, M.~W., Treu, T., Bolton, A.~S., {et~al.} 2010, \apj, 724, 511

\bibitem[{Baltz {et~al.}(2009)Baltz, Marshall, \& Oguri}]{Baltz2009}
Baltz, E.~A., Marshall, P., \& Oguri, M. 2009, JCAP, 2009, 015

\bibitem[{Barkana \& Loeb(2000)}]{Barkana1999}
Barkana, R., \& Loeb, A. 2000, \apj, 531, 613

\bibitem[{Barone-Nugent {et~al.}(2015)Barone-Nugent, Wyithe, Trenti, Treu,
  Oesch, Bouwens, Illingworth, \& Schmidt}]{Barone-Nugent2015}
Barone-Nugent, R.~L., Wyithe, J. S.~B., Trenti, M., {et~al.} 2015,
  arXiv:1502.03887

\bibitem[{Barone-Nugent {et~al.}(2013)Barone-Nugent, Wyithe, Trenti, Treu,
  Oesch, Bradley, \& Schmidt}]{Barone-Nugent2013a}
---. 2013, arXiv:1303.6109

\bibitem[{{Behroozi} \& {Silk}(2015)}]{Behroozi2015}
{Behroozi}, P.~S., \& {Silk}, J. 2015, \apj, 799, 32

\bibitem[{Belli {et~al.}(2014{\natexlab{a}})Belli, Newman, \&
  Ellis}]{Belli2014}
Belli, S., Newman, A.~B., \& Ellis, R.~S. 2014{\natexlab{a}}, \apj, 783, 117

\bibitem[{Belli {et~al.}(2014{\natexlab{b}})Belli, Newman, Ellis, \&
  Konidaris}]{Belli2014a}
Belli, S., Newman, A.~B., Ellis, R.~S., \& Konidaris, N.~P. 2014{\natexlab{b}},
  \apj, 788, L29

\bibitem[{Benitez {et~al.}(2004)Benitez, Ford, Bouwens, Menanteau, Blakeslee,
  Gronwall, Illingworth, Meurer, Broadhurst, Clampin, Franx, Hartig, Magee,
  Sirianni, Ardila, Bartko, Brown, Burrows, Cheng, Cross, Feldman, Golimowski,
  Infante, Kimble, Krist, Lesser, Levay, Martel, Miley, Postman, Rosati,
  Sparks, Tran, Tsvetanov, White, \& Zheng}]{Benitez2004}
Benitez, N., Ford, H., Bouwens, R.~J., {et~al.} 2004, \apjs, 150, 1

\bibitem[{{Bezanson} {et~al.}(2015){Bezanson}, {Franx}, \& {van
  Dokkum}}]{Bezanson2015}
{Bezanson}, R., {Franx}, M., \& {van Dokkum}, P.~G. 2015, \apj, 799, 148

\bibitem[{Bezanson {et~al.}(2012)Bezanson, van Dokkum, \& Franx}]{Bezanson2012}
Bezanson, R., van Dokkum, P., \& Franx, M. 2012, \apj, 760, 62

\bibitem[{Bezanson {et~al.}(2013)Bezanson, van Dokkum, van~de Sande, Franx,
  Leja, \& Kriek}]{Bezanson2013}
Bezanson, R., van Dokkum, P.~G., van~de Sande, J., {et~al.} 2013, \apj, 779,
  L21

\bibitem[{Bezanson {et~al.}(2011)Bezanson, van Dokkum, Franx, Brammer,
  Brinchmann, Kriek, Labb\'{e}, Quadri, Rix, van~de Sande, Whitaker, \&
  Williams}]{Bezanson2011}
Bezanson, R., van Dokkum, P.~G., Franx, M., {et~al.} 2011, \apj, 737, L31

\bibitem[{Blaizot {et~al.}(2005)Blaizot, Wadadekar, Guiderdoni, Colombi,
  Bertin, Bouchet, Devriendt, \& Hatton}]{Blaizot2005}
Blaizot, J., Wadadekar, Y., Guiderdoni, B., {et~al.} 2005, \mnras, 360, 159

\bibitem[{{Blandford} \& {Narayan}(1992)}]{Blandford1992}
{Blandford}, R.~D., \& {Narayan}, R. 1992, \araa, 30, 311

\bibitem[{Bouwens {et~al.}(2007)Bouwens, Illingworth, Franx, \&
  Ford}]{Bouwens2007}
Bouwens, R.~J., Illingworth, G.~D., Franx, M., \& Ford, H. 2007, \apj, 670, 928

\bibitem[{Bouwens {et~al.}(2011)Bouwens, Illingworth, Oesch, Labb\'{e}, Trenti,
  van Dokkum, Franx, Stiavelli, Carollo, Magee, \& Gonzalez}]{Bouwens2011}
Bouwens, R.~J., Illingworth, G.~D., Oesch, P.~A., {et~al.} 2011, \apj, 737, 90

\bibitem[{{Bouwens} {et~al.}(2013){Bouwens}, {Oesch}, {Illingworth},
  {Labb{\'e}}, {van Dokkum}, {Brammer}, {Magee}, {Spitler}, {Franx}, {Smit},
  {Trenti}, {Gonzalez}, \& {Carollo}}]{Bouwens2013}
{Bouwens}, R.~J., {Oesch}, P.~A., {Illingworth}, G.~D., {et~al.} 2013, \apjl,
  765, L16

\bibitem[{Bouwens {et~al.}(2014)Bouwens, Illingworth, Oesch, Trenti, Labbe',
  Bradley, Carollo, van Dokkum, Gonzalez, Holwerda, Franx, Spitler, Smit, \&
  Magee}]{Bouwens2014}
Bouwens, R.~J., Illingworth, G.~D., Oesch, P.~A., {et~al.} 2014,
  arXiv:1403.4295

\bibitem[{Bowler {et~al.}(2014{\natexlab{a}})Bowler, Dunlop, Mclure, Mccracken,
  Taniguchi, Fynbo, Le, \& Observatory}]{Bowler2014b}
Bowler, R. A.~A., Dunlop, J.~S., Mclure, R.~J., {et~al.} 2014{\natexlab{a}},
  arXiv:1411.2976

\bibitem[{Bowler {et~al.}(2014{\natexlab{b}})Bowler, Dunlop, McLure, Rogers,
  McCracken, Milvang-Jensen, Furusawa, Fynbo, Taniguchi, Afonso, Bremer, \& {Le
  Fevre}}]{Bowler2014a}
Bowler, R. A.~A., Dunlop, J.~S., McLure, R.~J., {et~al.} 2014{\natexlab{b}},
  \mnras, 440, 2810

\bibitem[{Bradley {et~al.}(2012)Bradley, Trenti, Oesch, Stiavelli, Treu,
  Bouwens, Shull, Holwerda, \& Pirzkal}]{Bradley2012}
Bradley, L.~D., Trenti, M., Oesch, P.~A., {et~al.} 2012, \apj, 760, 108

\bibitem[{{Bradley} {et~al.}(2014){Bradley}, {Zitrin}, {Coe}, {Bouwens},
  {Postman}, {Balestra}, {Grillo}, {Monna}, {Rosati}, {Seitz}, {Host}, {Lemze},
  {Moustakas}, {Moustakas}, {Shu}, {Zheng}, {Broadhurst}, {Carrasco}, {Jouvel},
  {Koekemoer}, {Medezinski}, {Meneghetti}, {Nonino}, {Smit}, {Umetsu},
  {Bartelmann}, {Ben{\'{\i}}tez}, {Donahue}, {Ford}, {Infante}, {Jimenez-Teja},
  {Kelson}, {Lahav}, {Maoz}, {Melchior}, {Merten}, \& {Molino}}]{Bradley2014}
{Bradley}, L.~D., {Zitrin}, A., {Coe}, D., {et~al.} 2014, \apj, 792, 76

\bibitem[{Bundy {et~al.}(2005)Bundy, Ellis, \& Conselice}]{Bundy2005}
Bundy, K., Ellis, R.~S., \& Conselice, C.~J. 2005, \apj, 625, 621

\bibitem[{{Cai} {et~al.}(2014){Cai}, {Lapi}, {Bressan}, {De Zotti}, {Negrello},
  \& {Danese}}]{Cai2014}
{Cai}, Z.-Y., {Lapi}, A., {Bressan}, A., {et~al.} 2014, \apj, 785, 65

\bibitem[{Cappellari {et~al.}(2006)Cappellari, Bacon, Bureau, Damen, Davies,
  {De Zeeuw}, Emsellem, Falcon-Barroso, Krajnovic, Kuntschner, McDermid,
  Peletier, Sarzi, {Van Den Bosch}, \& {Van De Ven}}]{Cappellari2006}
Cappellari, M., Bacon, R., Bureau, M., {et~al.} 2006, \mnras, 366, 1126

\bibitem[{Chae(2010)}]{Chae2010}
Chae, K.-H. 2010, \mnras, 402, 2031

\bibitem[{{Charlot} \& {Fall}(2000)}]{Charlot2000}
{Charlot}, S., \& {Fall}, S.~M. 2000, \apj, 539, 718

\bibitem[{Choi {et~al.}(2007)Choi, Park, \& Vogeley}]{Choi2007}
Choi, Y., Park, C., \& Vogeley, M.~S. 2007, \apj, 658, 884

\bibitem[{Coe {et~al.}(2006)Coe, Ben\'{\i}tez, S\'{a}nchez, Jee, Bouwens, \&
  Ford}]{Coe2006}
Coe, D., Ben\'{\i}tez, N., S\'{a}nchez, S.~F., {et~al.} 2006, \aj, 132, 926

\bibitem[{{Coe} {et~al.}(2015){Coe}, {Bradley}, \& {Zitrin}}]{Coe2015}
{Coe}, D., {Bradley}, L., \& {Zitrin}, A. 2015, \apj, 800, 84

\bibitem[{Coles \& Jones(1991)}]{Coles1991}
Coles, P., \& Jones, B. 1991, \mnras, 248, 1

\bibitem[{Collett {et~al.}(2013)Collett, Marshall, Auger, Hilbert, Suyu,
  Greene, Treu, Fassnacht, Koopmans, Brada\v{c}, \& Blandford}]{Collett2013}
Collett, T.~E., Marshall, P.~J., Auger, M.~W., {et~al.} 2013, \mnras, 432, 679

\bibitem[{Cooray(2005)}]{Cooray2005}
Cooray, A. 2005, \mnras, 364, 303

\bibitem[{Ellis {et~al.}(2001)Ellis, Santos, Kneib, \& Kuijken}]{Ellis2001}
Ellis, R., Santos, M.~R., Kneib, J.-P., \& Kuijken, K. 2001, \apj, 560, L119

\bibitem[{Faber \& Jackson(1976)}]{Faber1976}
Faber, S.~M., \& Jackson, R.~E. 1976, \apj, 204, 668

\bibitem[{Fassnacht {et~al.}(2004)Fassnacht, Moustakas, Casertano, Ferguson,
  Lucas, \& Park}]{Fassnacht2004}
Fassnacht, C.~D., Moustakas, L.~A., Casertano, S., {et~al.} 2004, \apj, 600,
  L155

\bibitem[{{Fialkov} \& {Loeb}(2015)}]{Fialkov2015}
{Fialkov}, A., \& {Loeb}, A. 2015, arXiv:1502.03141

\bibitem[{{Finkelstein} {et~al.}(2013){Finkelstein}, {Papovich}, {Dickinson},
  {Song}, {Tilvi}, {Koekemoer}, {Finkelstein}, {Mobasher}, {Ferguson},
  {Giavalisco}, {Reddy}, {Ashby}, {Dekel}, {Fazio}, {Fontana}, {Grogin},
  {Huang}, {Kocevski}, {Rafelski}, {Weiner}, \& {Willner}}]{Finkelstein2013}
{Finkelstein}, S.~L., {Papovich}, C., {Dickinson}, M., {et~al.} 2013, \nat,
  502, 524

\bibitem[{Finkelstein {et~al.}(2014)Finkelstein, {Ryan Jr.}, Papovich,
  Dickinson, Song, Somerville, Ferguson, Salmon, Giavalisco, Koekemoer, Ashby,
  Behroozi, Castellano, Dunlop, Faber, Fazio, Fontana, Grogin, Hathi, Jaacks,
  Kocevski, Livermore, McLure, Merlin, Mobasher, Newman, Rafelski, Tilvi, \&
  Willner}]{Finkelstein2014}
Finkelstein, S.~L., {Ryan Jr.}, R.~E., Papovich, C., {et~al.} 2014,
  arXiv:1410.5439

\bibitem[{Franx {et~al.}(1997)Franx, Illingworth, Kelson, van Dokkum, \&
  Tran}]{Franx1997}
Franx, M., Illingworth, G.~D., Kelson, D.~D., van Dokkum, P.~G., \& Tran, K.-V.
  1997, \apj, 486, L75

\bibitem[{Gardner {et~al.}(2006)Gardner, Mather, Clampin, Doyon, Greenhouse,
  Hammel, Hutchings, Jakobsen, Lilly, Long, Lunine, Mccaughrean, Mountain,
  Nella, Rieke, Rieke, Rix, Smith, Sonneborn, Stiavelli, Stockman, Windhorst,
  \& Wright}]{JWST_SSR}
Gardner, J., Mather, J., Clampin, M., {et~al.} 2006, Space Science Reviews,
  123, 485

\bibitem[{Greene {et~al.}(2013)Greene, Suyu, Treu, Hilbert, Auger, Collett,
  Marshall, Fassnacht, Blandford, Brada\v{c}, \& Koopmans}]{Greene2013}
Greene, Z.~S., Suyu, S.~H., Treu, T., {et~al.} 2013, \apj, 768, 39

\bibitem[{{Grogin} {et~al.}(2011){Grogin}, {Kocevski}, {Faber}, {Ferguson},
  {Koekemoer}, {Riess}, {Acquaviva}, {Alexander}, {Almaini}, {Ashby}, {Barden},
  {Bell}, {Bournaud}, {Brown}, {Caputi}, {Casertano}, {Cassata}, {Castellano},
  {Challis}, {Chary}, {Cheung}, {Cirasuolo}, {Conselice}, {Roshan Cooray},
  {Croton}, {Daddi}, {Dahlen}, {Dav{\'e}}, {de Mello}, {Dekel}, {Dickinson},
  {Dolch}, {Donley}, {Dunlop}, {Dutton}, {Elbaz}, {Fazio}, {Filippenko},
  {Finkelstein}, {Fontana}, {Gardner}, {Garnavich}, {Gawiser}, {Giavalisco},
  {Grazian}, {Guo}, {Hathi}, {H{\"a}ussler}, {Hopkins}, {Huang}, {Huang},
  {Jha}, {Kartaltepe}, {Kirshner}, {Koo}, {Lai}, {Lee}, {Li}, {Lotz}, {Lucas},
  {Madau}, {McCarthy}, {McGrath}, {McIntosh}, {McLure}, {Mobasher},
  {Moustakas}, {Mozena}, {Nandra}, {Newman}, {Niemi}, {Noeske}, {Papovich},
  {Pentericci}, {Pope}, {Primack}, {Rajan}, {Ravindranath}, {Reddy}, {Renzini},
  {Rix}, {Robaina}, {Rodney}, {Rosario}, {Rosati}, {Salimbeni}, {Scarlata},
  {Siana}, {Simard}, {Smidt}, {Somerville}, {Spinrad}, {Straughn}, {Strolger},
  {Telford}, {Teplitz}, {Trump}, {van der Wel}, {Villforth}, {Wechsler},
  {Weiner}, {Wiklind}, {Wild}, {Wilson}, {Wuyts}, {Yan}, \& {Yun}}]{Grogin2011}
{Grogin}, N.~A., {Kocevski}, D.~D., {Faber}, S.~M., {et~al.} 2011, \\apjs, 197,
  35

\bibitem[{Guo {et~al.}(2011)Guo, White, Boylan-Kolchin, {De Lucia}, Kauffmann,
  Lemson, Li, Springel, \& Weinmann}]{Guo2011}
Guo, Q., White, S., Boylan-Kolchin, M., {et~al.} 2011, \mnras, 413, 101

\bibitem[{Hathi {et~al.}(2012)Hathi, Mobasher, Capak, Wang, \&
  Ferguson}]{Hathi2012}
Hathi, N.~P., Mobasher, B., Capak, P., Wang, W.-h., \& Ferguson, H.~C. 2012,
  \apj, 757, 43

\bibitem[{{Hayes} {et~al.}(2012){Hayes}, {Laporte}, {Pell{\'o}}, {Schaerer}, \&
  {Le Borgne}}]{Hayes2012}
{Hayes}, M., {Laporte}, N., {Pell{\'o}}, R., {Schaerer}, D., \& {Le Borgne},
  J.-F. 2012, \mnras, 425, L19

\bibitem[{Henriques {et~al.}(2012)Henriques, White, Lemson, Thomas, Guo,
  Marleau, \& Overzier}]{Henriques2012}
Henriques, B. M.~B., White, S. D.~M., Lemson, G., {et~al.} 2012, \mnras, 421,
  2904

\bibitem[{Hilbert {et~al.}(2009)Hilbert, Hartlap, White, \&
  Schneider}]{Hilbert2009}
Hilbert, S., Hartlap, J., White, S. D.~M., \& Schneider, P. 2009, \aap, 499, 31

\bibitem[{Hilbert {et~al.}(2007)Hilbert, White, Hartlap, \&
  Schneider}]{Hilbert2007}
Hilbert, S., White, S. D.~M., Hartlap, J., \& Schneider, P. 2007, \mnras, 382,
  121

\bibitem[{Kelly {et~al.}(2008)Kelly, Fan, \& Vestergaard}]{Kelly2008}
Kelly, B.~C., Fan, X., \& Vestergaard, M. 2008, \apj, 682, 874

\bibitem[{Kitzbichler \& White(2007)}]{Kitzbichler2007}
Kitzbichler, M.~G., \& White, S. D.~M. 2007, \mnras, 376, 2

\bibitem[{{Koekemoer} {et~al.}(2011){Koekemoer}, {Faber}, {Ferguson}, {Grogin},
  {Kocevski}, {Koo}, {Lai}, {Lotz}, {Lucas}, {McGrath}, {Ogaz}, {Rajan},
  {Riess}, {Rodney}, {Strolger}, {Casertano}, {Castellano}, {Dahlen},
  {Dickinson}, {Dolch}, {Fontana}, {Giavalisco}, {Grazian}, {Guo}, {Hathi},
  {Huang}, {van der Wel}, {Yan}, {Acquaviva}, {Alexander}, {Almaini}, {Ashby},
  {Barden}, {Bell}, {Bournaud}, {Brown}, {Caputi}, {Cassata}, {Challis},
  {Chary}, {Cheung}, {Cirasuolo}, {Conselice}, {Roshan Cooray}, {Croton},
  {Daddi}, {Dav{\'e}}, {de Mello}, {de Ravel}, {Dekel}, {Donley}, {Dunlop},
  {Dutton}, {Elbaz}, {Fazio}, {Filippenko}, {Finkelstein}, {Frazer}, {Gardner},
  {Garnavich}, {Gawiser}, {Gruetzbauch}, {Hartley}, {H{\"a}ussler},
  {Herrington}, {Hopkins}, {Huang}, {Jha}, {Johnson}, {Kartaltepe},
  {Khostovan}, {Kirshner}, {Lani}, {Lee}, {Li}, {Madau}, {McCarthy},
  {McIntosh}, {McLure}, {McPartland}, {Mobasher}, {Moreira}, {Mortlock},
  {Moustakas}, {Mozena}, {Nandra}, {Newman}, {Nielsen}, {Niemi}, {Noeske},
  {Papovich}, {Pentericci}, {Pope}, {Primack}, {Ravindranath}, {Reddy},
  {Renzini}, {Rix}, {Robaina}, {Rosario}, {Rosati}, {Salimbeni}, {Scarlata},
  {Siana}, {Simard}, {Smidt}, {Snyder}, {Somerville}, {Spinrad}, {Straughn},
  {Telford}, {Teplitz}, {Trump}, {Vargas}, {Villforth}, {Wagner}, {Wandro},
  {Wechsler}, {Weiner}, {Wiklind}, {Wild}, {Wilson}, {Wuyts}, \&
  {Yun}}]{Koekemoer2011}
{Koekemoer}, A.~M., {Faber}, S.~M., {Ferguson}, H.~C., {et~al.} 2011, \apjs,
  197, 36

\bibitem[{{Kriek} {et~al.}(2006){Kriek}, {van Dokkum}, {Franx}, {Quadri},
  {Gawiser}, {Herrera}, {Illingworth}, {Labb{\'e}}, {Lira}, {Marchesini},
  {Rix}, {Rudnick}, {Taylor}, {Toft}, {Urry}, \& {Wuyts}}]{Kriek2006}
{Kriek}, M., {van Dokkum}, P.~G., {Franx}, M., {et~al.} 2006, \apjl, 649, L71

\bibitem[{Laureijs {et~al.}(2011)Laureijs, Amiaux, Arduini, Augueres,
  Brinchmann, Cole, Cropper, Dabin, Duvet, Ealet, {et~al.}}]{Laureijs2011}
Laureijs, R., Amiaux, J., Arduini, S., {et~al.} 2011, arXiv:1110.3193

\bibitem[{Maraston(2005)}]{Maraston2005}
Maraston, C. 2005, \mnras, 362, 799

\bibitem[{McCully {et~al.}(2014)McCully, Keeton, Wong, \&
  Zabludoff}]{McCully2014}
McCully, C., Keeton, C.~R., Wong, K.~C., \& Zabludoff, A.~I. 2014, \mnras, 443,
  3631

\bibitem[{McLure {et~al.}(2013)McLure, Dunlop, Bowler, Curtis-Lake, Schenker,
  Ellis, Robertson, Koekemoer, Rogers, Ono, Ouchi, Charlot, Wild, Stark,
  Furlanetto, Cirasuolo, \& Targett}]{McLure2013}
McLure, R.~J., Dunlop, J.~S., Bowler, R. A.~A., {et~al.} 2013, \mnras, 432,
  2696

\bibitem[{Mu\~{n}oz(2012)}]{Munoz2012}
Mu\~{n}oz, J.~A. 2012, JCAP, 2012, 015

\bibitem[{Muzzin {et~al.}(2013)Muzzin, Marchesini, Stefanon, Franx, McCracken,
  Milvang-Jensen, Dunlop, Fynbo, Brammer, Labb\'{e}, \& van
  Dokkum}]{Muzzin2013}
Muzzin, A., Marchesini, D., Stefanon, M., {et~al.} 2013, \apj, 777, 18

\bibitem[{Newman {et~al.}(2010)Newman, Ellis, Treu, \& Bundy}]{Newman2010}
Newman, A.~B., Ellis, R.~S., Treu, T., \& Bundy, K. 2010, \apj, 717, L103

\bibitem[{Oesch {et~al.}(2012)Oesch, Bouwens, Illingworth, Gonzalez, Trenti,
  van Dokkum, Franx, Labb\'{e}, Carollo, \& Magee}]{Oesch2012}
Oesch, P.~A., Bouwens, R.~J., Illingworth, G.~D., {et~al.} 2012, \apj, 759, 135

\bibitem[{Ono {et~al.}(2012)Ono, Ouchi, Mobasher, Dickinson, Penner, Shimasaku,
  Weiner, Kartaltepe, Nakajima, Nayyeri, Stern, Kashikawa, \&
  Spinrad}]{Ono2012}
Ono, Y., Ouchi, M., Mobasher, B., {et~al.} 2012, \apj, 744, 83

\bibitem[{Parzen(1962)}]{parzen1962}
Parzen, E. 1962, The Annals of Mathematical Statistics, 33, 1065

\bibitem[{Pei(1995)}]{Pei1995}
Pei, Y.~C. 1995, \apj, 440, 485

\bibitem[{Robertson {et~al.}(2013)Robertson, Furlanetto, Schneider, Charlot,
  Ellis, Stark, McLure, Dunlop, Koekemoer, Schenker, Ouchi, Ono, Curtis-Lake,
  Rogers, Bowler, \& Cirasuolo}]{Robertson2013}
Robertson, B.~E., Furlanetto, S.~R., Schneider, E., {et~al.} 2013, \apj, 768,
  71

\bibitem[{Rosenblatt(1956)}]{rosenblatt1956}
Rosenblatt, M. 1956, The Annals of Mathematical Statistics, 27, 832

\bibitem[{Schechter(1976)}]{Schechter1976}
Schechter, P. 1976, \apj, 203, 297

\bibitem[{Schenker {et~al.}(2013)Schenker, Robertson, Ellis, Ono, McLure,
  Dunlop, Koekemoer, Bowler, Ouchi, Curtis-Lake, Rogers, Schneider, Charlot,
  Stark, Furlanetto, \& Cirasuolo}]{Schenker2013}
Schenker, M.~A., Robertson, B.~E., Ellis, R.~S., {et~al.} 2013, \apj, 768, 196

\bibitem[{Schmidt {et~al.}(2014{\natexlab{a}})Schmidt, Treu, Trenti, Bradley,
  Kelly, Oesch, Holwerda, Shull, \& Stiavelli}]{Schmidt2014}
Schmidt, K.~B., Treu, T., Trenti, M., {et~al.} 2014{\natexlab{a}}, \apj, 786,
  57

\bibitem[{Schmidt {et~al.}(2014{\natexlab{b}})Schmidt, Treu, Brammer,
  Brada\v{c}, Wang, Dijkstra, Dressler, Fontana, Gavazzi, Henry, Hoag, Jones,
  Kelly, Malkan, Mason, Pentericci, Poggianti, Stiavelli, Trenti, von~der
  Linden, \& Vulcani}]{Schmidt2014a}
Schmidt, K.~B., Treu, T., Brammer, G.~B., {et~al.} 2014{\natexlab{b}}, \apj,
  782, L36

\bibitem[{{Schneider} {et~al.}(1992){Schneider}, {Ehlers}, \&
  {Falco}}]{Schneider1992}
{Schneider}, P., {Ehlers}, J., \& {Falco}, E.~E. 1992, {Gravitational Lenses}
  (Springer)

\bibitem[{{Schneider} {et~al.}(2006){Schneider}, {Kochanek}, \&
  {Wambsganss}}]{SaasFee}
{Schneider}, P., {Kochanek}, C.~S., \& {Wambsganss}, J. 2006, {Gravitational
  Lensing: Strong, Weak and Micro} (Springer)

\bibitem[{Sheth {et~al.}(2003)Sheth, Bernardi, Schechter, Burles, Eisenstein,
  Finkbeiner, Frieman, Lupton, Schlegel, Subbarao, Shimasaku, Bahcall,
  Brinkmann, \& Ivezi\'{c}}]{Sheth2003}
Sheth, R.~K., Bernardi, M., Schechter, P.~L., {et~al.} 2003, \apj, 594, 225

\bibitem[{{Somerville} {et~al.}(2012){Somerville}, {Gilmore}, {Primack}, \&
  {Dom{\'{\i}}nguez}}]{Somerville2012}
{Somerville}, R.~S., {Gilmore}, R.~C., {Primack}, J.~R., \& {Dom{\'{\i}}nguez},
  A. 2012, \mnras, 423, 1992

\bibitem[{Spergel {et~al.}(2013)Spergel, Gehrels, Breckinridge, Donahue,
  Dressler, Gaudi, Greene, Guyon, Hirata, Kalirai, {et~al.}}]{Spergel2013}
Spergel, D., Gehrels, N., Breckinridge, J., {et~al.} 2013, arXiv:1305.5425

\bibitem[{Springel {et~al.}(2005)Springel, White, Jenkins, Frenk, Yoshida, Gao,
  Navarro, Thacker, Croton, Helly, Peacock, Cole, Thomas, Couchman, Evrard,
  Colberg, \& Pearce}]{Springel2005}
Springel, V., White, S. D.~M., Jenkins, A., {et~al.} 2005, Nature, 435, 629

\bibitem[{Suyu {et~al.}(2010)Suyu, Marshall, Auger, Hilbert, Blandford,
  Koopmans, Fassnacht, \& Treu}]{Suyu2010}
Suyu, S.~H., Marshall, P.~J., Auger, M.~W., {et~al.} 2010, \apj, 711, 201

\bibitem[{Toft {et~al.}(2012)Toft, Gallazzi, Zirm, Wold, Zibetti, Grillo, \&
  Man}]{Toft2012}
Toft, S., Gallazzi, A., Zirm, A., {et~al.} 2012, \apj, 754, 3

\bibitem[{Trenti \& Stiavelli(2008)}]{Trenti2008}
Trenti, M., \& Stiavelli, M. 2008, \apj, 676, 767

\bibitem[{Trenti {et~al.}(2011)Trenti, Bradley, Stiavelli, Oesch, Treu,
  Bouwens, Shull, MacKenty, Carollo, \& Illingworth}]{Trenti2011}
Trenti, M., Bradley, L.~D., Stiavelli, M., {et~al.} 2011, \apj, 727, L39

\bibitem[{Trenti {et~al.}(2012)Trenti, Bradley, Stiavelli, Shull, Oesch,
  Bouwens, Mu\~{n}oz, Romano-Diaz, Treu, Shlosman, \& Carollo}]{Trenti2012}
---. 2012, \apj, 746, 55

\bibitem[{Treu(2010)}]{Treu2010}
Treu, T. 2010, \araa, 48, 87

\bibitem[{Treu {et~al.}(2005)Treu, Ellis, Liao, \& van Dokkum}]{Treu2005}
Treu, T., Ellis, R.~S., Liao, T.~X., \& van Dokkum, P.~G. 2005, \apj, 622, L5

\bibitem[{Turner {et~al.}(1984)Turner, Ostriker, \& {Gott, J. R.}}]{Turner1984}
Turner, E.~L., Ostriker, J.~P., \& {Gott, J. R.}, I. 1984, \apj, 284, 1

\bibitem[{Vale \& Ostriker(2004)}]{Vale2004}
Vale, a., \& Ostriker, J.~P. 2004, \mnras, 353, 189

\bibitem[{van~de Sande {et~al.}(2014)van~de Sande, Kriek, Franx, Bezanson, \&
  van Dokkum}]{VandeSande2014}
van~de Sande, J., Kriek, M., Franx, M., Bezanson, R., \& van Dokkum, P.~G.
  2014, \apj, 793, L31

\bibitem[{van~de Sande {et~al.}(2013)van~de Sande, Kriek, Franx, van Dokkum,
  Bezanson, Bouwens, Quadri, Rix, \& Skelton}]{VandeSande2013}
van~de Sande, J., Kriek, M., Franx, M., {et~al.} 2013, \apj, 771, 85

\bibitem[{van~der Burg {et~al.}(2010)van~der Burg, Hildebrandt, \&
  Erben}]{VanderBurg2010}
van~der Burg, R. F.~J., Hildebrandt, H., \& Erben, T. 2010, \aap, 523, A74

\bibitem[{van~der Wel {et~al.}(2008)van~der Wel, Holden, Zirm, Franx, Rettura,
  Illingworth, \& Ford}]{VanderWel2008}
van~der Wel, A., Holden, B.~P., Zirm, A.~W., {et~al.} 2008, \apj, 688, 48

\bibitem[{{van Dokkum} {et~al.}(2009){van Dokkum}, {Kriek}, \&
  {Franx}}]{vanDokkum2009}
{van Dokkum}, P.~G., {Kriek}, M., \& {Franx}, M. 2009, \nat, 460, 717

\bibitem[{{Windhorst} {et~al.}(2006){Windhorst}, {Cohen}, {Jansen},
  {Conselice}, \& {Yan}}]{Windhorst2006}
{Windhorst}, R.~A., {Cohen}, S.~H., {Jansen}, R.~A., {Conselice}, C., \& {Yan},
  H. 2006, NAR, 50, 113

\bibitem[{{Windhorst} {et~al.}(2011){Windhorst}, {Cohen}, {Hathi}, {McCarthy},
  {Ryan}, {Yan}, {Baldry}, {Driver}, {Frogel}, {Hill}, {Kelvin}, {Koekemoer},
  {Mechtley}, {O'Connell}, {Robotham}, {Rutkowski}, {Seibert}, {Straughn},
  {Tuffs}, {Balick}, {Bond}, {Bushouse}, {Calzetti}, {Crockett}, {Disney},
  {Dopita}, {Hall}, {Holtzman}, {Kaviraj}, {Kimble}, {MacKenty}, {Mutchler},
  {Paresce}, {Saha}, {Silk}, {Trauger}, {Walker}, {Whitmore}, \&
  {Young}}]{Windhorst2011}
{Windhorst}, R.~A., {Cohen}, S.~H., {Hathi}, N.~P., {et~al.} 2011, \apjs, 193,
  27

\bibitem[{Wyithe {et~al.}(2011)Wyithe, Yan, Windhorst, \& Mao}]{Wyithe2011}
Wyithe, J., Yan, H., Windhorst, R.~R., \& Mao, S. 2011, Nature, 469, 181

\bibitem[{Wyithe {et~al.}(2001)Wyithe, Turner, \& Spergel}]{Wyithe2000}
Wyithe, J. S.~B., Turner, E.~L., \& Spergel, D.~N. 2001, \apj, 555, 504

\bibitem[{Yan {et~al.}(2011)Yan, Yan, Zamojski, Windhorst, McCarthy, Fan,
  R\"{o}ttgering, Koekemoer, Robertson, Dav\'{e}, \& Cai}]{Yan2011}
Yan, H., Yan, L., Zamojski, M.~A., {et~al.} 2011, \apj, 728, L22

\bibitem[{{Zitrin} {et~al.}(2015){Zitrin}, {Fabris}, {Merten}, {Melchior},
  {Meneghetti}, {Koekemoer}, {Coe}, {Maturi}, {Bartelmann}, {Postman},
  {Umetsu}, {Seidel}, {Sendra}, {Broadhurst}, {Balestra}, {Biviano}, {Grillo},
  {Mercurio}, {Nonino}, {Rosati}, {Bradley}, {Carrasco}, {Donahue}, {Ford},
  {Frye}, \& {Moustakas}}]{Zitrin2015}
{Zitrin}, A., {Fabris}, A., {Merten}, J., {et~al.} 2015, \apj, 801, 44

\end{thebibliography}
%======================================================================
\end{document}